\documentclass[11pt]{article}

\usepackage[english]{babel}

\usepackage[letterpaper,top=2cm,bottom=2cm,left=3cm,right=3cm,marginparwidth=1.75cm]{geometry}
\usepackage{tikz}
\usetikzlibrary{arrows,positioning}
\definecolor{boxcolor}{HTML}{B9DCFF}
\usetikzlibrary{arrows.meta}
\usepackage{subcaption}
\usepackage{float}
\usepackage{multirow}
\usepackage{booktabs}
\usepackage{pgfplots}
\pgfplotsset{compat=1.18} 
\usepackage{amsfonts}
\usepackage{amsmath, graphicx, epsfig,psfrag, verbatim}
\usepackage{eucal,amssymb,amsthm,graphicx, amsfonts, latexsym,xcolor}
\usepackage[utf8]{inputenc}

\usepackage{footnote}
\makesavenoteenv{tabular}
\makesavenoteenv{table}

\usepackage[normalem]{ulem}
\newcommand\redout{\bgroup\markoverwith
{\textcolor{red}{\rule[0.5ex]{2pt}{0.8pt}}}\ULon}

\usepackage[colorlinks=true, allcolors=blue]{hyperref}

\newtheorem{prop}{Proposition}

\title{SIRS epidemics with individual heterogeneity of immunity waning}
\author{Mohamed El Khalifi and Tom Britton}

\begin{document}
\maketitle

\begin{abstract}
In the current paper we analyse an extended SIRS epidemic model in which immunity at the individual level wanes gradually at exponential rate, but where the waning rate may differ between individuals, for instance as an effect of differences in immune systems. The model also includes vaccination schemes aimed to reach and maintain herd immunity. We consider both the \emph{informed} situation where the individual waning parameters are known, thus allowing selection of vaccinees being based on both time since last vaccination as well as on the individual waning rate, and the more likely \emph{uninformed} situation where individual waning parameters are unobserved, thus only allowing vaccination schemes to depend on time since last vaccination. The optimal vaccination policies for both the informed and uniformed heterogeneous situation are derived and compared with the homogeneous waning model (meaning all individuals have the same immunity waning rate),  as well as to the classic SIRS model where immunity at the individual level drops from complete immunity to complete susceptibility in one leap. It is shown that the classic SIRS model requires least vaccines, followed by the SIRS with homogeneous gradual waning, followed by the informed situation for the model with heterogeneous gradual waning. The situation requiring most vaccines for herd immunity is the most likely scenario, that immunity wanes gradually with unobserved individual heterogeneity. For parameter values chosen to mimic COVID-19 and assuming perfect initial immunity and cumulative immunity of 12 months, the classic homogeneous SIRS epidemic suggests that vaccinating individuals every 15 months is sufficient to reach and maintain herd immunity, whereas the uninformed case for exponential waning with rate heterogeneity corresponding to a coefficient of variation being 0.5, requires that individuals instead need to be vaccinated every 4.4 months.

\end{abstract}

\section{Introduction}

Among other things, the COVID-19 pandemic showed that immunity waning as well as immunity escape for new virus strains play important roles when designing vaccination schemes to reduce and ultimately stop the spreading of an epidemic. In the current paper the focus lies on immunity waning for a fixed and specific strain and we thus study an epidemic model for an infectious disease where immunity, both from vaccination as well as natural infection, wanes gradually and monotonically following an exponential mode \cite{wheatley2021evolution}. 

The classic SIRS model \cite{hethcote1976qualitative,hethcote1978immunization} is the first model to consider immunity waning, and in this model population immunity decays gradually, but at the individual level each individual is either fully immune or fully susceptible. In the last few years this assumption has been relaxed (e.g. \cite{reluga2008backward,martcheva2015introduction,forien2022stochastic,khalifi2022extending}) thus allowing for gradual waning of immunity also at the individual level, resulting in individuals having different immunity levels, defined either discretely or continuously. These models still assume that immunity wanes in a similar fashion for all individuals, most often defined by a waning rate $\omega$ common for all individuals. For these models it has been shown that such gradual immunity waning requires more frequent vaccination to reach and maintain herd immunity as compared to the classic SIRS model which assumes one single jump from fully immune to fully susceptible  (having the same average cumulative immunity) \cite{khalifi2022extending}. 

Empirical measurements of antibodies however suggest large individual differences in antibody decay between individuals \cite{fabiani2022effectiveness,shrotri2021spike,widge2021durability,perez2022modeling} thus suggesting different waning rates between individuals. In the present paper we therefore extend a model with homogeneous gradual immunity waning to a situation where the waning rate may differ between individuals. The general situation, where waning rates of individuals are drawn independently from some general random distribution is complicated to analyse, so here we focus on the situation where there are two types of waning rates $\omega_1$ and $\omega_2$  with population frequencies $p$ and $1-p$ respectively. We compare the heterogeneous situation with the homogeneous case having the same cumulative immunity ($=1/\omega$) and we quantify the amount of heterogeneity by the coefficient of variation of the immunity distribution.

This paper is structured as follows. In the next section we present the SIRS models with heterogeneity under both situations: sudden loss and continuous waning of immunity. In Section \ref{SIRkS_sec} we formulate the SIR\textsuperscript{$(k)$}S model with heterogeneity. In Section \ref{vacc_sec}, we introduce vaccination into the model by taking into account the effect of the available information on individuals immunity. To illustrate the results for our models, in Section \ref{res_sec},  we compare the long term prevalence and the optimal vaccination schemes under parameter values mimicking the COVID-19 pandemic. We conclude the paper in Section \ref{dis_sec} with a discussion and draw some perspectives.

\section{Models}\label{mod_sec}
First, we define a model where immune individuals lose their immunity at once. Next we modify the model to allow for gradual (exponential) waning of individual immunity. For both models we divide the population into two immunity waning classes with waning rates $\omega_1$ and $\omega_2$ with fractions $p$ and $1-p$, allowing for a certain degree of heterogeneity. We compare the homogeneous case with cumulative immunity $1/\omega$ to the heterogeneous case with $1/\omega_1= (1-\alpha)(1/\omega)$ and $1/\omega_2= (1+\alpha \frac{p}{1-p} )(1/\omega)$ for some $\alpha$ ($0\le \alpha\le 1$) so that the cumulative immunity is set to $1/\omega$. The coefficient of variation of the immunity distribution is given by $\sigma=\alpha\sqrt{p/(1-p)}$ which from now on is used as heterogeneity parameter rather than $\alpha$. Hence, we have
\begin{align*}
    \frac{1}{\omega_1} = \frac{1}{\omega} \left( 1-\sigma\sqrt{\frac{1-p}{p}}\right) \mbox{ and }\, \frac{1}{\omega_2} = \frac{1}{\omega} \left( 1+\sigma\sqrt{\frac{p}{1-p}}\right).
\end{align*}
\subsection{The SIRS model with heterogeneity}
The model that we consider in this section, also taking births and deaths into account, is defined as follows. Let $l\in\{1,2\}$ be the index of the immunity waning classes and denote $s_l(t),i_l(t)$, and $r_l(t)$ the community fractions of susceptible, infectious, and recovered $l$-individuals at time $t$, respectively. The model parameters are as defined in Table \ref{table:Params-values}.
\begin{table}[ht]
	\renewcommand{\arraystretch}{1.5}
	\centering
	\caption{Model Parameters and interpretation.}
	\label{table:Params-values}
	\begin{tabular}[t]{cc}
		\toprule
		Parameter &Description\\
		\midrule
		$ \mu $ &Birth and death rate\\
		$ \beta $ &Effective infection rate\\
		$ \gamma $ &Recovery rate \\
		$ \omega $ &Immunity waning rate \\
		$ \sigma $ &Coefficient of variation of immunity distribution \\
		\bottomrule
	\end{tabular}
\end{table}
Then, the differential equations for the SIRS model with heterogeneity are given by
\begin{align}\label{ODE1}
		s_l'(t) &=p_l\,\mu - \beta s_l(t) \left(i_1(t)+i_2(t)\right) - \mu s_l(t) + \omega_l r_l(t),\nonumber\\
		i_l'(t) &= \beta s_l(t)   \left(i_1(t)+i_2(t)\right) - (\gamma+\mu ) i_l(t),\\
		r_l'(t) &= \gamma i_l(t) - \left( \mu+ \omega_l\right) r_l(t),\nonumber
\end{align}
with $l\in\{1,2\}$ (so $s_l(t)+i_l(t)+r_l(t)=p_l$), $p_1=p$ and $p_2=1-p$. We define the basic reproduction number to be $R_0 = \frac{\beta}{\gamma+\mu}$ representing the average number of new infections generated by an infectious person in an entirely susceptible population. We have the following standard result for our model.
\begin{prop}\label{exist_uniq_het_SIRS}
    The solution to Eq.\  \eqref{ODE1} has a unique endemic equilibrium if and only if $R_0>1$. 
\end{prop}
When the endemic equilibrium exist, the endemic level is given by the sum of the constant fractions of infectives $\hat i_1$ and $\hat i_2$ in the type-1 and type-2 communities respectively.
We also have the following result regarding the dependence of the endemic level on the population heterogeneity. Recall the $p$ is the community fraction having lower immunity and hence higher waning rate $\omega_1$.
\begin{prop}\label{end level increasing}
    Assume that $R_0>1$ and $ 1/2\le p<1$. Then, the endemic level is an increasing function of the coefficient of variation $\sigma$ on $[0,\sqrt{p/(1-p)})$.
\end{prop}
The proofs of the Propositions \ref{exist_uniq_het_SIRS} and \ref{end level increasing} are given in the Appendix \ref{append: proofs of props}. Although the monotonicity in Proposition \ref{end level increasing} is only proved for a fraction $p$ satisfying $ 1/2\le p<1$, numerical simulations suggest that the endemic level is also increasing in $\sigma$ on $[0,\sqrt{p/(1-p)})$ for any $0< p< 1/2.$

\subsection[The SIRinftyS model with heterogeneity]{The SIR\textsuperscript{$(\infty)$}S model with heterogeneity}
When immunity wanes continuously and following an exponential decay, the recovered equation could be modelled using a PDE evolving in calendar time $t$ and time-since-recovery $a$, and the model equations become
\begin{align}\label{ODE-PDE1}
		s_l'(t) &=p_l\mu - \beta s_l(t) (i_1(t)+i_2(t)) - \mu s_l(t),\nonumber\\
		i_l'(t) &= \beta \left( s_l(t)  +  \int_0^{\infty}\, \left( 1-e^{-\omega_l a}\right) \,  r_l(t,a)\, da \right) \, (i_1(t)+i_2(t)) - (\gamma+\mu ) i_l(t),\\
		\dfrac{\partial r_l(t,a)}{\partial t} + \dfrac{\partial r_l(t,a)}{\partial a} &= - \beta  \left( 1-e^{-\omega_l a}\right)\, r_l(t,a) (i_1(t)+i_2(t)) - \mu r_l(t,a),\quad a>0,\nonumber
\end{align}
with the boundary condition $r_l(t,0) = \gamma i_l(t),$ $l\in\{1,2\}$ where $p_1=p$ and $p_2=1-p$.\\
We call the model \eqref{ODE-PDE1} the SIR\textsuperscript{$(\infty)$}S model with heterogeneity as it can be seen as the $k$-limit of the heterogeneous SIR\textsuperscript{$(k)$}S model where immunity drops in $k$ steps, $1/k$ each time \cite{khalifi2022extending}. We refer to \cite{khalifi2022extending} for more details on the construction (see also Section \ref{SIRkS_sec}). 
We have the following expected result for the model \eqref{ODE-PDE1}.
\begin{prop}\label{exist_uniq_het_contin} 
~~
\begin{itemize}
    \item Assume that $R_0\leq1$. Then, the solution to Eq. \eqref{ODE-PDE1} converges to the disease-free equilibrium.
    \item Assume that $R_0>1$. Then, Eq. \eqref{ODE-PDE1} has a unique endemic equilibrium.
\end{itemize}
\end{prop}
The proof of the Proposition \ref{exist_uniq_het_contin} is given in Appendix \ref{App_exist_uniq_het_contin}.

\section{SIR\textsuperscript{$(k)$}S model with heterogeneity}\label{SIRkS_sec}

Similarly to the approach in \cite{khalifi2022extending}, we approximate the SIR\textsuperscript{$(\infty)$}S model \eqref{ODE-PDE1} by a system of ODEs allowing immunity to wane in $k$ steps for some large value of $k$. 

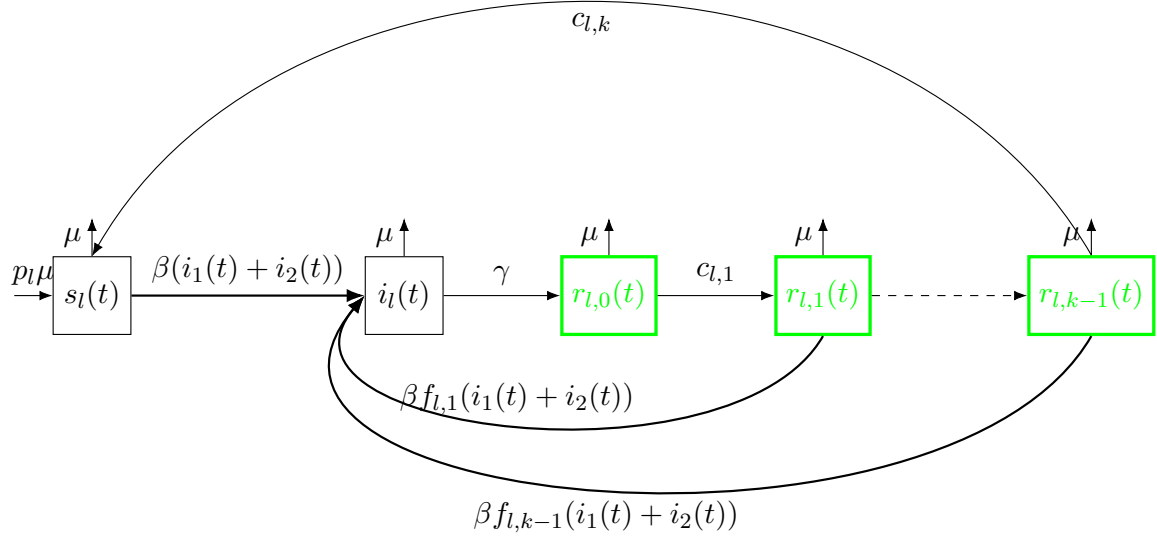
\begin{figure}
\resizebox{\textwidth}{!}{
\begin{tikzpicture}[node distance=1cm, auto,
    >=Latex, 
    every node/.append style={align=center},
    int/.style={draw, minimum size=1cm}]
\centering
   \node [int] (S1)             {$s_l(t)$};
   \node (b1) [left of=S1, coordinate] {S1}; 
   \node [int, right= 3cm of S1] (I1) {$i_l(t)$};
   \node [color=green,very thick,int, right= 1.5cm of I1] (R10) {$r_{l,0}(t)$};
   \node [color=green,very thick,int, right= 1.5cm of R10] (R11) {$r_{l,1}(t)$};
   \node [color=green,very thick,int, right= 2cm of R11] (R1k1) {$r_{l,k-1}(t)$};

\node (bb11) [above of=S1, coordinate] {S1};
\node (bb12) [above of=I1, coordinate] {I1};
\node (bb13) [above of=R10, coordinate] {R10};
\node (bb14) [above of=R11, coordinate] {R11};
\node (bb15) [above of=R1k1, coordinate] {R1k1};
   
   \draw [dashed,->] (R11) -- node [above] {}  (R1k1);
   
   \coordinate[right=of I1] (out);
   \path[->] (S1) edge node {$\mu$} (bb11);
   \path[->] (I1) edge node {$\mu$} (bb12);
   \path[->] (R10) edge node {$\mu$} (bb13);
   \path[->] (R11) edge node {$\mu$} (bb14);
   \path[->] (R1k1) edge node {$\mu$} (bb15);

   \path[->] (b1) edge node {$p_l\mu$} (S1);
   \path[->] (S1) edge[thick] node {$\beta (i_1(t)+i_2(t))$} (I1)
             (I1) edge node {$\gamma$} (R10)
             (R10) edge node {$c_{l,1}$} (R11)
               (R1k1.south)  edge[out=-120, in=-130, thick] node[below] {$\beta f_{l,k-1} (i_1(t)+i_2(t))$} (I1.west)
               (R1k1.north)  edge[out=120, in=60] node {$c_{l,k}$} (S1.north)
               (R11.south) edge[out=-120, in=-130, thick] node[above] {$\beta f_{l,1} (i_1(t)+i_2(t))$} (I1.west);
\end{tikzpicture}
}
\caption{Diagram of the SIR\textsuperscript{(k)}S epidemic model in the $l$-type individuals, $l=1,2$.}
\label{fig:SIRkS}
\end{figure}

We now describe how this reduction of immunity in $k$ small steps down to no immunity, each step having a high rate to drop to the next level. This can be done in several ways still reaching the same continuous limit as $k\to\infty$ and it is convenient to choose different choices for different constructions why we define the general construction. The most important thing is however that for large $k$ immunity jumps in many small steps, each having a high jump rate.  Let $  \{ r_{l,j}(t)\}_{j=1}^{k-1}$, $l\in \{1,2\}$  be the fractions of recovered individuals, at time t, with the immunity levels $  \{ 1- f_{l,j}\}_{j=1}^{k-1} $ (or susceptibility levels $  \{ f_{l,j}\}_{j=1}^{k-1} $), $l\in \{1,2\}$, and $  \{ c_{l,j}\}_{j=1}^k $,  $l\in \{1,2\}$ be the rates at which recovered individuals lose immunity portions through the $k$ steps. Since $k$ is fixed and typically large, we drop it from the notation. The resulting model equations are given by
\begin{equation}\label{SIRkS}
	\begin{array}{lll}
		s_l'(t) &=& p_l\mu - \beta s_l(t) \, (i_1(t)+i_2(t)) + c_{l,k} r_{l,k-1}(t) - \mu s_l(t),\\
		i_l'(t) &=& \beta \big( s_l(t) +  \sum\limits_{j=1}^{k-1} f_{l,j} r_{l,j}(t) \big) \, (i_1(t)+i_2(t)) - (\gamma+\mu) i_l(t),\\
		r_{l,0}'(t) &=& \gamma i_l(t) - (c_{l,1}+\mu) r_{l,0}(t),\\
		r_{l,j}'(t) &=& c_{l,j} r_{l,j-1}(t)  - \beta f_{l,j} r_{l,j}(t)\, (i_1(t)+i_2(t)) - (c_{l,j+1}+\mu) r_{l,j}(t), 
	\end{array}
\end{equation}
for $j=1,\cdots,k-1$ and $l\in \{1,2\}$ where $p_1=p$ and $p_2=1-p$. We call this model the SIR\textsuperscript{$(k)$}S model with heterogeneity. See Fig. \ref{fig:SIRkS} for a transition scheme and the Appendix \ref{model_formulation} for a derivation of the immunity jumps and the transition rates (see also \cite{khalifi2022extending}).
\section{SIR\textsuperscript{$(k)$}S model with heterogeneity and vaccination}\label{vacc_sec}

When immunity wanes over time, it is important to allow the vaccination strategies to depend on time since last vaccination and vaccines should not be uniformly distributed.  Hence the vaccination rate in the $j$'th susceptibility class of an $l$-type individuals might depend on both $l$ and $j$. Let $\eta_{l,j}$ to denote this vaccination rate. A vaccination strategy is hence specified by these rates $\{ \eta_{l,j}\}$, many rates often being 0 since strategies would often be defined by vaccinating once immunity drops to a certain level. Then, the SIR\textsuperscript{$(k)$}S model with vaccination is given by the following equations
\begin{equation}\label{SIRkS_vacc}
	\begin{array}{lll}
		s_l'(t) &=& p_l\mu - \beta s_l(t) \, (i_1(t)+i_2(t)) + c_{l,k} r_{l,k-1}(t) - \left( \mu + \eta_{l,k} \right) s_l(t),\\
		i_l'(t) &=& \beta \big( s_l(t) +  \sum\limits_{j=1}^{k-1} f_{l,j} r_{l,j}(t) \big) \, (i_1(t)+i_2(t)) - (\gamma+\mu) i_l(t),\\
		r_{l,0}'(t) &=& \eta_{l,k} s_l(t) + \sum\limits_{j=1}^{k-1}\eta_{l,j} r_{l,j}(t) + \gamma i_l(t) - (c_{l,1}+\mu) r_{l,0}(t),\\
		r_{l,j}'(t) &=& c_{l,j} r_{l,j-1}(t)  - \beta f_{l,j} r_{l,j}(t)\, (i_1(t)+i_2(t)) - (c_{l,j+1}+\mu+\eta_{l,j}) r_{l,j}(t), 
	\end{array}
\end{equation}
for $j=1,\cdots,k-1$ and $l\in \{1,2\}$ where $p_1=p$ and $p_2=1-p$. 

What is the best, or optimal, vaccination strategy differs depending on amount of available information. In the situation where no information is available, a potential strategy could be to randomly vaccinate in all non-infectious classes, including individuals with partial immunity. Here we distinguish between two situations: both individual time since last vaccination and waning rates are known (informed situation), and only time since last vaccination is known
(uninformed situation).

\subsection{Informed situation}
When the individual waning rate and time since last vaccination for all individuals are known, vaccines will be administrated to those with faster waning at different frequency compared to individuals with slower waning. In this situation, we assume that the susceptibility levels $ \{ f_{l,j}\}_{j=1}^{k-1} $ are the same for both types and $f_{l,j}=\frac{j}{k}, j=1,\cdots,k-1$. Hence the transition rates $ \{ c_{l,j}\}_{j=1}^k $,  $l\in \{1,2\}$ (depend on $l$) are such that the average cumulative immunity equals $1/\omega_1$ and $1/\omega_2$ respectively (see Appendix \ref{model_formulation}). For any $l\in \{1,2\}$, we denote by $\eta_{l,j}$ the vaccination rate in the class $r_{l,j}$ for $j=1,\cdots,k-1$, and by $\eta_{l,k}$ the vaccination rate of fully susceptible individuals  $s_l$. The resulting model equations are given by
\begin{equation}\label{SIRkS vacc inf}
	\begin{array}{lll}
		s_l'(t) &=& p_l\mu - \beta s_l(t) \, (i_1(t)+i_2(t)) + c_{l,k} r_{l,k-1}(t) - \left( \mu + \eta_{l,k} \right) s_l(t),\\
		i_l'(t) &=& \beta \big( s_l(t) +  \sum\limits_{j=1}^{k-1}\frac{j}{k} r_{l,j}(t) \big) \, (i_1(t)+i_2(t)) - (\gamma+\mu) i_l(t),\\
		r_{l,0}'(t) &=& \eta_{l,k} s_l(t) + \sum\limits_{j=1}^{k-1}\eta_{l,j} r_{l,j}(t) +  \gamma i_l(t) - (c_{l,1}+\mu) r_{l,0}(t),\\
		r_{l,j}'(t) &=& c_{l,j} r_{l,j-1}(t)  - \beta \frac{j}{k}r_{l,j}(t)\, (i_1(t)+i_2(t)) - (c_{l,j+1}+\mu+\eta_{l,j}) r_{l,j}(t) , 
	\end{array}
\end{equation}
for $j=1,\cdots,k-1$, and $l\in \{1,2\}$ where $p_1=p$ and $p_2=1-p$.

\begin{figure}[ht]
\centering
    \begin{subfigure}[b]{0.48\linewidth}        
        \centering
        \includegraphics[width=\linewidth]{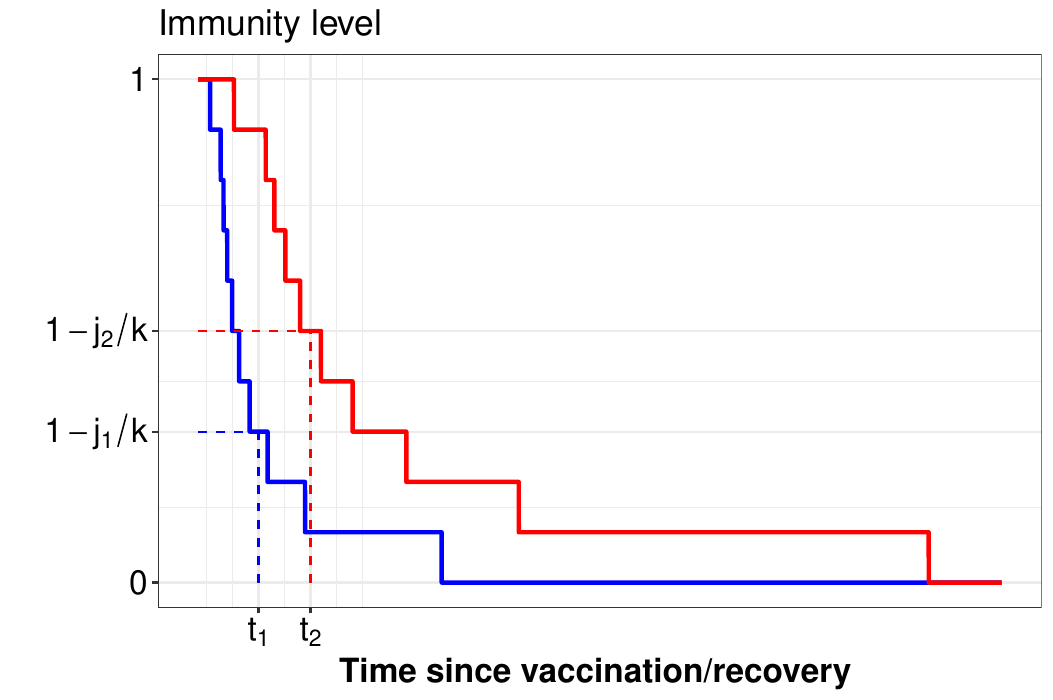}
        \caption{}
        \label{fig:steps_het}
    \end{subfigure}
    \begin{subfigure}[b]{0.48\linewidth}        
        \centering
        \includegraphics[width=\linewidth]{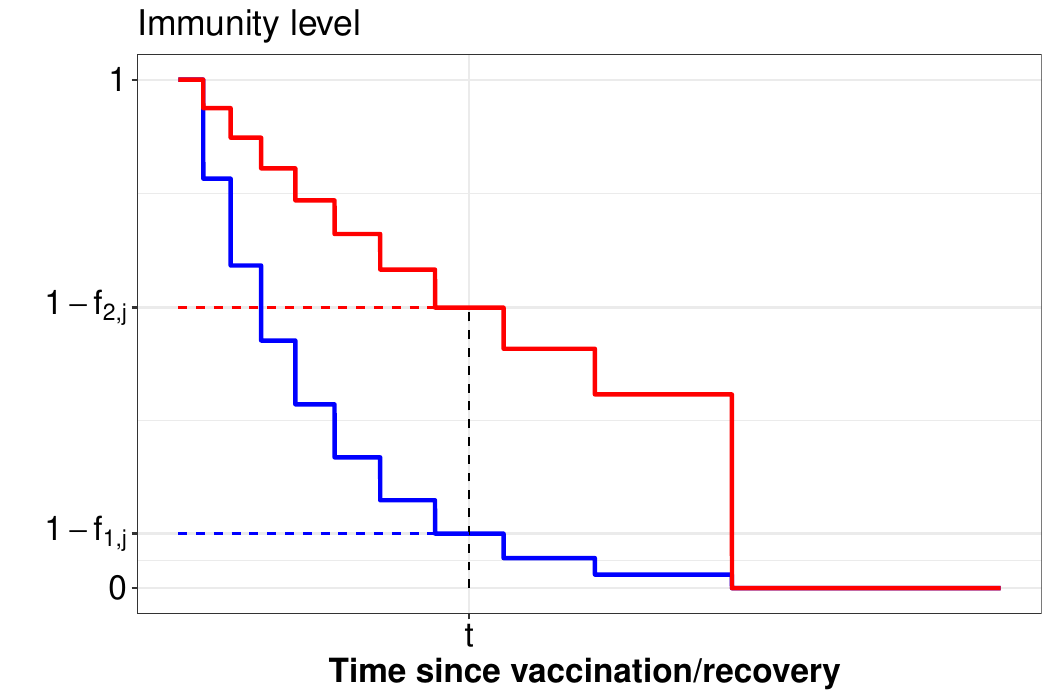}
        \caption{}
        \label{fig:steps_het_uninf}
    \end{subfigure}
    \caption{Examples showing the vaccination strategy when immunity wanes in $k=10$ steps in (a) the informed situation (b) the uninformed situation.}
    \label{fig:steps}
\end{figure}

Here the immunity class of each individual is known. The best vaccination scheme is then to vaccinate 1-individuals once they have lost $j_1$ steps of immunity and 2-individuals once they have lost $j_2$ immunity steps, for some values of $j_1$ and $j_2$ (see Fig. \ref{fig:steps_het}). This corresponds to vaccinating the two types of individuals at (possible different) fixed times, $t_1$ and $t_2$ respectively, since last vaccination (or infection). Clearly, the smaller $j_1$ and $j_2$ the more vaccines are needed, and the optimal relation between $j_1$ and $j_2$ will depend on the immunity waning rates $\omega_1$ and $\omega_2$.

\subsection{Uninformed situation}
Here we consider the more realistic situation where only time since last recovery/vaccination is known. In this situation, we let the transition rates $ \{ c_{l,j}\}_{j=1}^k,l\in\{1,2\} $ to be independent on $l$, and then the susceptibilities $ \{ f_{l,j}\}_{j=1}^k $ are no longer the same for $l\in \{1,2\}$ (see Appendix \ref{model_formulation}). In addition, for both $l\in \{1,2\}$ we let $\eta_{j}$ to be the vaccination rate in the class $r_{l,j}$ for $j=1,\cdots,k-1$, and $\eta_{k}$ to be the vaccination rate of fully susceptible individuals $s_l$ (the vaccination rate must be the same for both types in the uninformed situation). The resulting model equations are given by
\begin{equation}\label{SIRkS vacc uninf}
	\begin{array}{lll}
		s_l'(t) &=& p_l\mu - \beta s_l(t) \, (i_1(t)+i_2(t)) + c_{k} r_{l,k-1}(t) - \left( \mu + \eta_{k} \right) s_l(t),\\
		i_l'(t) &=& \beta \big( s_l(t) +  \sum\limits_{j=1}^{k-1} f_{l,j} r_{l,j}(t) \big) \, (i_1(t)+i_2(t)) - (\gamma+\mu) i_l(t),\\
		r_{l,0}'(t) &=& \eta_{k} s_l(t) + \sum\limits_{j=1}^{k-1}\eta_{j} r_{l,j}(t) +  \gamma i_l(t) - (c_{1}+\mu) r_{l,0}(t),\\
		r_{l,j}'(t) &=& c_{j} r_{l,j-1}(t)  - \beta f_{l,j} r_{l,j}(t)\, (i_1(t)+i_2(t)) - (c_{j+1}+\mu) r_{l,j}(t) - \eta_{j} r_{l,j}(t) , 
	\end{array}
\end{equation}
for $j=1,\cdots,k-1$, and $l\in \{1,2\}$ where $p_1=p$ and $p_2=1-p$.

Here vaccination is the same for both types since they are unobserved, and all individuals are vaccinated (at the same time, $t$) once they have lost $j$ steps of immunity (see Fig. \ref{fig:steps_het_uninf}).


\subsection{Extending to imperfect (leaky) vaccines}
In the previous section, both infection and vaccination are assumed to initially confer perfect immunity. This could be relaxed by considering imperfect vaccines producing partial protection level to any vaccinated person. Although, these partial immunities could differ between the two subpopulations, we here consider a leaky vaccine conferring immunity $e$ to all vaccinated individuals in both subpopulations.

\subsection{Reproduction number and optimal vaccination}
Recall that $e$ is the protection level that vaccines are assumed to confer to any vaccinated individual. Each constant vaccination scheme gives rise to a disease free equilibrium $E_0 = \left(\hat{s}_{1},\hat{s}_{2}, \hat{r}_{1,0}, \hat{r}_{2,0}, \cdots, \hat{r}_{1,k-1}, \hat{r}_{2,k-1} \right)$ (see Appendix \ref{App_DFE}). The corresponding reproduction number $R_{v}$ is given by
\begin{align}
    R_{v} = R_0 \, \sum_{l=1}^2 \left(   \hat{s}_{l} + e\,\sum_{j=1}^{k-1} f_{l,j}^k \hat{r}_{l,j} \right).
\end{align}
In the expressions $E_0$ and $R_v$, we omit the dependence on the vaccination strategy (informed or uninformed) for the sake of convenience.
Within a certain class of vaccination schemes, the optimal is the one solving the following optimization problem
\begin{align}\label{criticalv}
\theta_c^{k}= \min\limits_{\boldsymbol{\eta}} \, \theta^k (\boldsymbol{\eta}) \quad
\mbox{subject to} \quad R_v \leq 1,\nonumber
\end{align}
where $\theta^k (\boldsymbol{\eta})$ is the vaccine usage given by
\begin{equation}
\theta^{k}(\boldsymbol{\eta})= \sum_{l=1}^2 \eta_{l,k}\hat{s}_l +  \sum\limits_{j=1}^{k-1} \eta_{l,j}\hat{r}_{l,j},
\end{equation}
for a $(2\times k)$-matrix of vaccination rates $\boldsymbol{\eta}$ within the class of possible vaccination schemes.

\subsubsection*{Informed optimal vaccination strategy}
Within each type it is always better to vaccinate less immune individuals compared to more immune individuals.
The optimal vaccination strategy in the informed situation is hence to vaccinate 1-individuals and 2-individuals as soon as their immunities drop below some levels $\iota_1 = 1-j_1/k$ and $\iota_2=1-j_2/k$, respectively, for some $j_1$ and $j_2$. For finite $k$ and by referring to model \eqref{SIRkS vacc inf}, this is equivalent to not vaccinate type-1 individuals in states $\left(r_{1,0},r_{1,1},\cdots,r_{1,j_1-1}\right)$ (and $\left(r_{2,0},r_{2,1},\cdots,r_{2,j_2-1}\right)$ for type-2 individuals) up to some $j_1, j_2\in\{1,\cdots,k\}$, to vaccinate in $r_{1,j_1}$ (resp. $r_{1,j_2}$) at some rate $\eta_{1,j_1}^\star$ (resp. $\eta_{1,j_2}^\star$), and to immediately vaccinate individuals leaving the state $r_{1,j_1}$ (resp. $r_{1,j_2}$). The states $(j_1,j_2)$ and the rates $(\eta_{1,j_1}^\star, \eta_{1,j_2}^\star)$ correspond to the minimal immunity levels and the minimal vaccination rates respectively, satisfying $R_v \leq 1$. For large $k$ this means we vaccinate type-1 and type-2 individuals once their immunities have dropped to $1-j_1/k$ and $1-j_2/k$ respectively. What are the optimal values of $j_1$ and $j_2$ for a given overall vaccination rate $\theta$ we solve numerically.

\subsubsection*{Uninformed optimal vaccination strategy} 
As only individual time since vaccination is known, the optimal vaccination strategy consists of vaccinating all individuals, irrespective of type, as soon as they reach some time $t$ since their last vaccination. The shorter $t$, the bigger the vaccine coverage $\theta_c^k$. For finite $k$ and by referring to model \eqref{SIRkS vacc uninf}, this is to not vaccinate up to some $j\in\{1,\cdots,k\}$, to vaccinate in both $r_{1,j}$ and $r_{2,j}$ at some rate $\eta_j^\star$, and immediately vaccinate individuals leaving the states $r_{1,j}$ and $r_{2,j}$. The state $j$ and the rate $\eta_j^\star$ correspond to the minimal immunity levels and the minimal vaccination rate respectively, satisfying $R_v \leq 1$ and this we also solve numerically. At time $t$, individuals immunities are different and equal to $\iota_{u,1}=1-f_{1,j}$ and $\iota_{u,2}=1-f_{2,j}$ for type-1 individuals and type-2 individuals respectively.
\newline
$~$
\newline
\textbf{Remark:} While both informed and uninformed vaccination strategies will be considered when immunity wanes gradually, only the informed situation is considered when immunity wanes in one jump. Indeed, introducing an uninformed vaccination in the SIRS model necessitates the change of the distribution of immunity duration and probably using a PDE model for the dynamics of vaccinated/recovered individuals, something which we do not consider in this paper.
\section{Results}\label{res_sec}

We now illustrate our results numerically, studying the effect of heterogeneity of the gradual waning, for parameter values consistent with Covid-19 (of course lacking many other features of reality). Our primary focus is to study the effect of heterogeneity measured by its coefficient of variation $\sigma$, but also to compare the informed situation, which assumes the individual heterogeneities to be known, to the uninformed case. We also compare our model to the classical homogeneous SIRS epidemic model as well as to the heterogeneous SIRS (loosing all immunity at once).

To illustrate how various waning assumptions affect disease prevalence and the vaccination frequency needed to avoid an outbreak to occur, we use the following parameter values. The life expectancy is set to $\mu^{-1}= 80$ years, the mean infectious period is set to $\gamma^{-1}= 0.02$ years (one week) and the average cumulative immunity is to $\omega^{-1}=1$ year. These parameter values are reasonable for several infectious diseases including COVID-19, influenza, common cold, etc \cite{byrne2020inferred,davies2020age,hall2022protection,cdc}. Although the waning rate $\omega$ could be estimated using for instance the antibody decay data \cite{goldberg2021waning,bobrovitz2023protective}, we do not attempt to do so here. The amount of waning heterogeneity is measured by the coefficient of variation $\sigma$ of immunity heterogeneity, which is always smaller than $\sqrt{p/(1-p)}$. We vary $\beta$ (or equivalently $R_0$) and $\sigma$ (often with $p=50\%$ fixed but sometimes also varying $p$). Using the different models we compare the endemic prevalence levels without vaccination, and the required amount of vaccines to reach a sustainable herd immunity.

\subsection{Endemic prevalence}
\begin{figure}[h]
\centering
    \begin{subfigure}[b]{0.48\linewidth}        
        \centering
        \includegraphics[width=\linewidth]{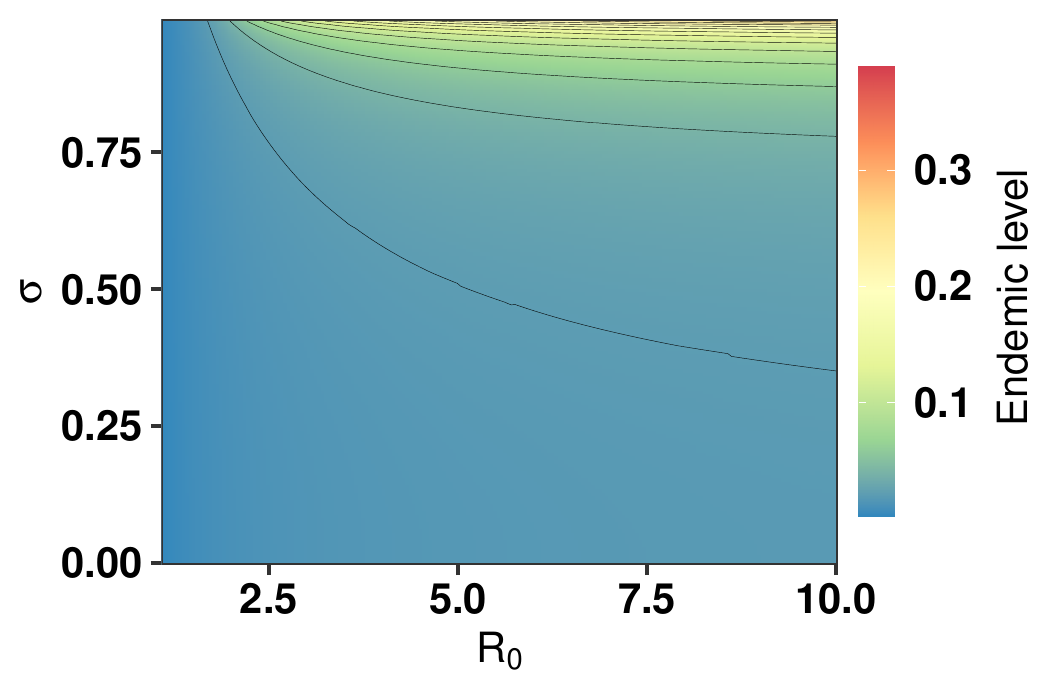}
        \caption{}
        \label{fig:endemic level hSIRS}
    \end{subfigure}
    \begin{subfigure}[b]{0.48\linewidth}        
        \centering
        \includegraphics[width=\linewidth]{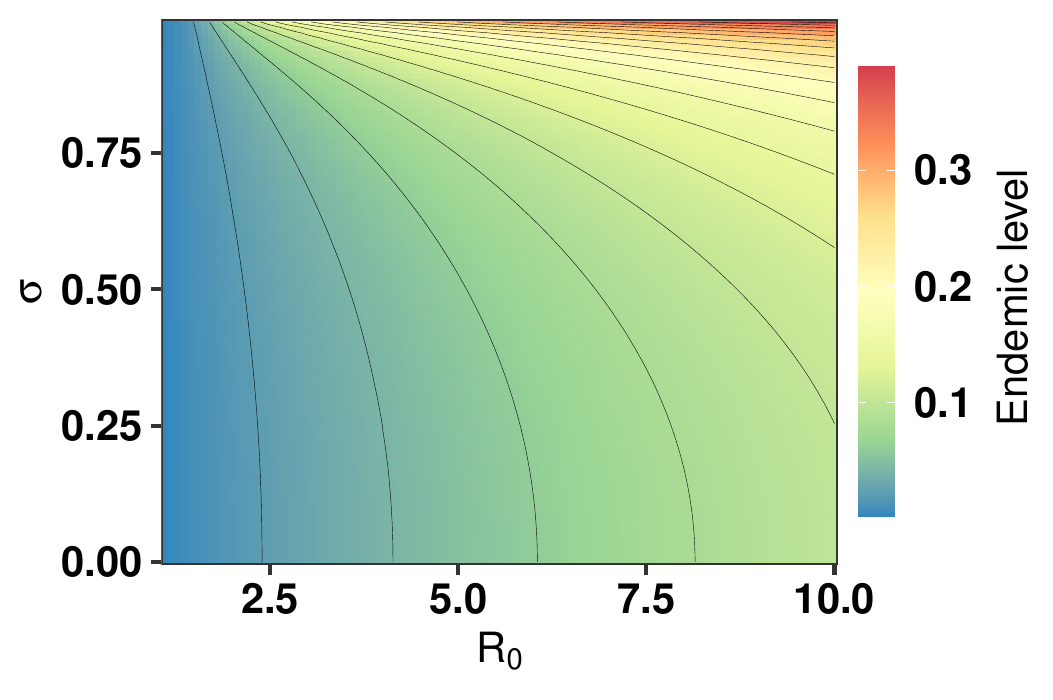}
        \caption{}
        \label{fig:endemic level hSIRkS}
    \end{subfigure}
    \caption{Heatmaps of the endemic level for different values of $R_0$ and $\sigma$ ($p=50\%$) in (a) the heterogeneous SIRS model and (b) the heterogeneous SIR\textsuperscript{$(\infty)$}S model. The value at the origin is 0 and the contour interval is $2\%$ of the population.}
    \label{fig:endemic levels}
\end{figure}

Fig. \ref{fig:endemic levels} shows heatmaps of the endemic level for the heterogeneous SIRS model \eqref{ODE1} and our new heterogeneous SIR\textsuperscript{$(\infty)$}S model \eqref{ODE-PDE1}, as functions of $R_0$ and $\sigma$ (without vaccination). It can be seen from Figs. \ref{fig:endemic level hSIRS}-\ref{fig:endemic level hSIRkS} that the long term prevalence is increasing in $R_0$ (as expected) but also in population heterogeneity $\sigma$. Moreover, Fig. \ref{fig:endemic levels p5-95_app} in the Appendix \ref{append: proofs of props} shows that the difference between homogeneous and heterogeneous populations increases with $p$ (the community fraction having the lower immunity, i.e.\ higher waning rate). The homogeneous models ($\sigma=0$) have the lowest endemic levels irrespective of the immunity waning mode (sudden or gradual loss).

\subsection{Optimal vaccination: Perfect vaccine}
\subsubsection{Possible vaccination times}
\begin{figure}[!htb]
\centering
    \begin{subfigure}[b]{0.6\linewidth}        
        \centering
        \includegraphics[width=\linewidth,height=6cm]{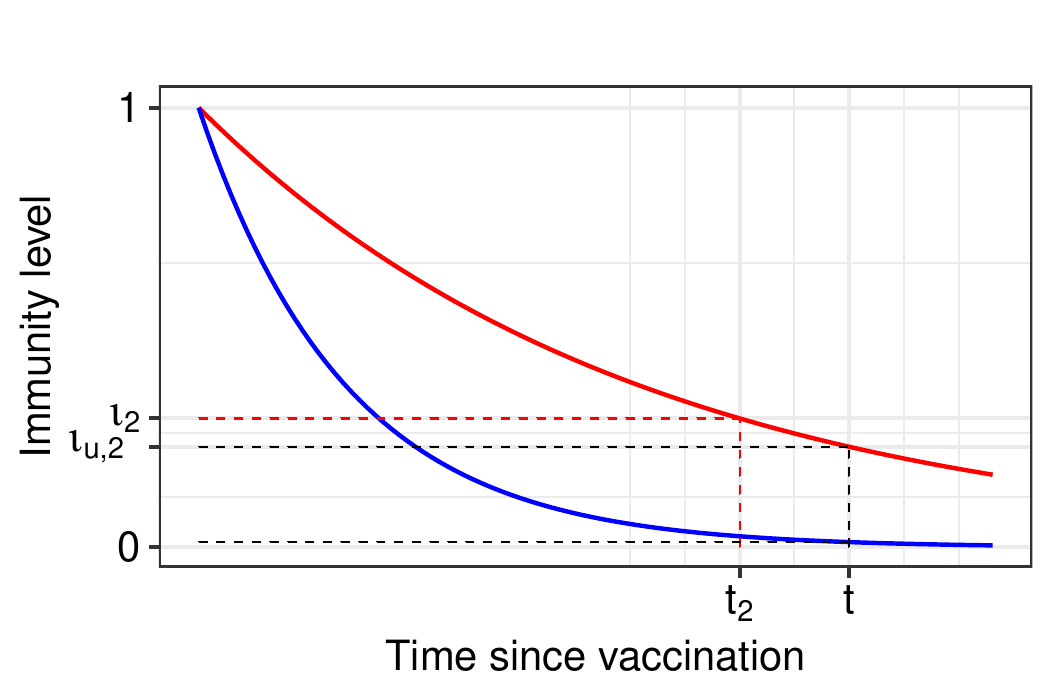}
        \caption{$R_0=1.6$.}
        \label{fig:crit_time_R_16}
    \end{subfigure}
    \begin{subfigure}[b]{0.6\linewidth}        
        \centering
        \includegraphics[width=\linewidth,height=6cm]{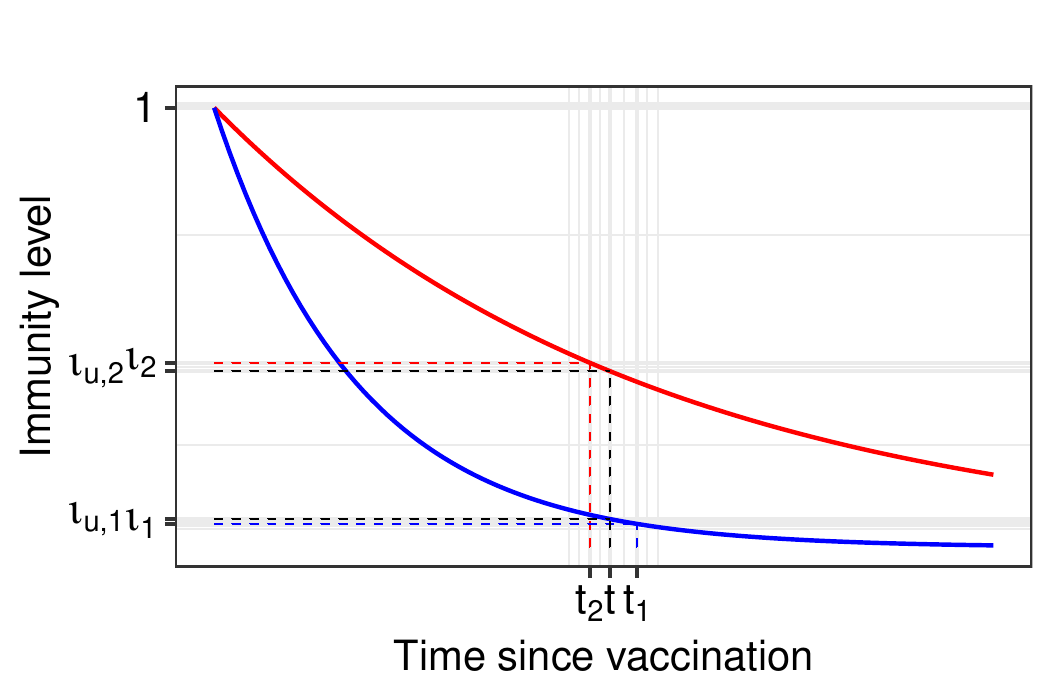}
        \caption{$R_0=2$.}
        \label{fig:crit_time_R_20}
    \end{subfigure}
    \begin{subfigure}[b]{0.6\linewidth}        
        \centering
        \includegraphics[width=\linewidth,height=6cm]{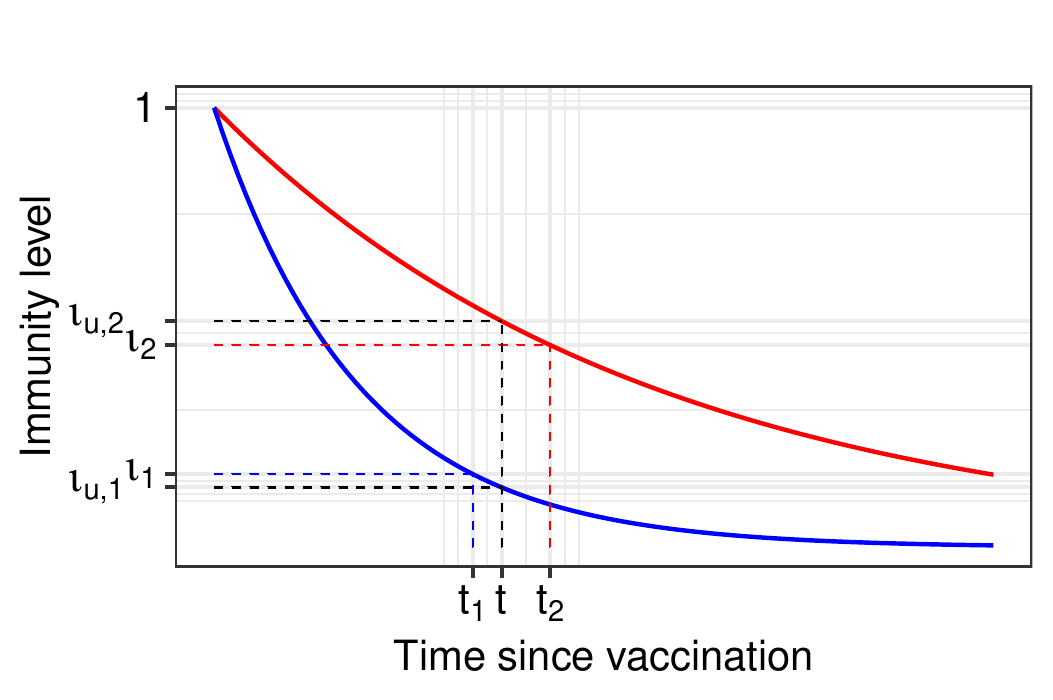}
        \caption{$R_0=2.4$.}
        \label{fig:crit_time_R_24}
    \end{subfigure}
\caption{Informed and the uninformed optimal vaccination times for different values of $R_0$ with $\sigma=0.5$ ($p=50\%$). Blue and Red solid curves are immunity waning functions of type-1 and type-2 individuals respectively. The best informed vaccination strategy is to vaccinate 1-individuals at time $t_1$ (with the immunity level $\iota_1$) and 2-individuals at time $t_2$ (with the immunity level $\iota_2$). The best uninformed vaccination strategy is to vaccinate everyone at time $t$, that is, 1-individuals and 2-individuals at the immunity levels $\iota_{u,1}$ and $\iota_{u,2}$ respectively.}
\label{fig:critical_vacc_times_R0}
\end{figure}

Fig. \ref{fig:critical_vacc_times_R0} shows the best vaccination strategies for the informed and uninformed situations.
The optimal strategy always vaccinate type 1 (with higher waning rate) at a lower immunity level compared to type 2 individuals (both informed and uninformed). In the informed situation it may even be optimal to only vaccinate type 2 individuals, e.g.\ when $R_0$ is small enough (Fig. \ref{fig:crit_time_R_16}) or if $\omega_1$ is very large so these individuals loose their immunity very quickly implying that there is not much gain in vaccinating them. However, the time (since vaccination) at which we vaccinate type 1 could be bigger or smaller than the time for type 2.  Plot \ref{fig:crit_time_R_16} shows that only type-2 need to be vaccinated, at time $t_2$, when $R_0=1.6$, but later on also type-1 have to be vaccinated, at time $t_1$, when $R_0$ increases to 2 as shown in Fig. \ref{fig:crit_time_R_20}. When $R_0=2.4$, the vaccination time $t_1$ of type-1 becomes smaller compared to $t_2$ for type-2 as seen in Fig. \ref{fig:crit_time_R_24}. In the uninformed situation all individuals need to be vaccinated after the same time $t$ since their last vaccination.

\subsubsection{Optimal vaccination scheme}
\begin{figure}[!htb]
\centering
    \begin{subfigure}[b]{0.49\linewidth}        
        \centering
        \includegraphics[width=\linewidth]{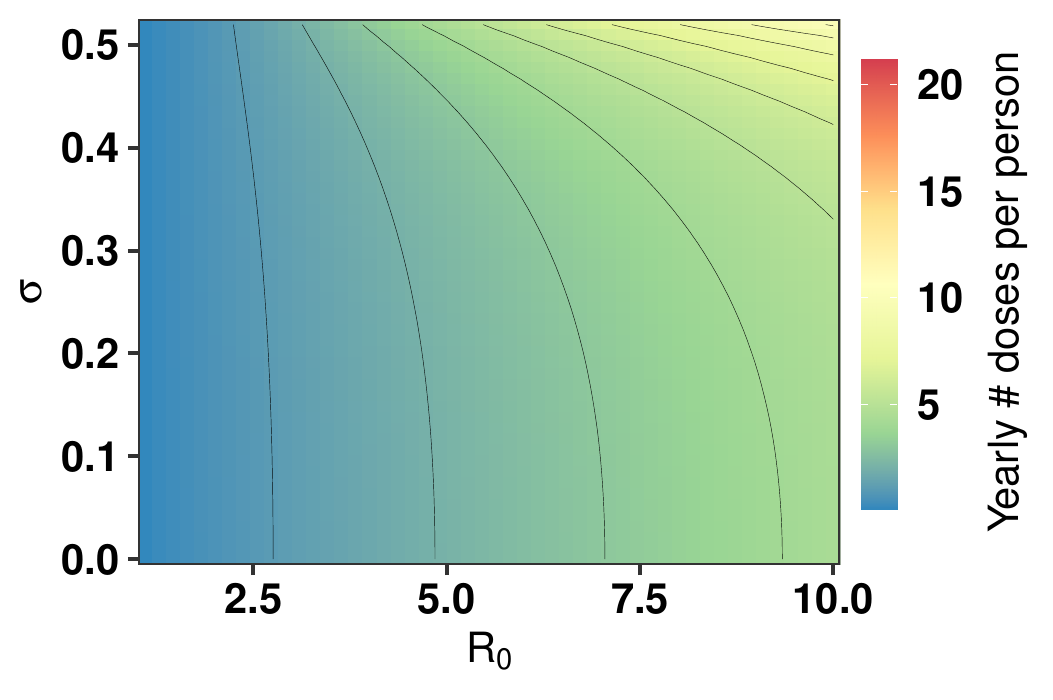}
        \caption{$p=25\%$.}
        \label{fig:criti ccine supply hSIRS}
    \end{subfigure}
    \begin{subfigure}[b]{0.49\linewidth}        
        \centering
        \includegraphics[width=\linewidth]{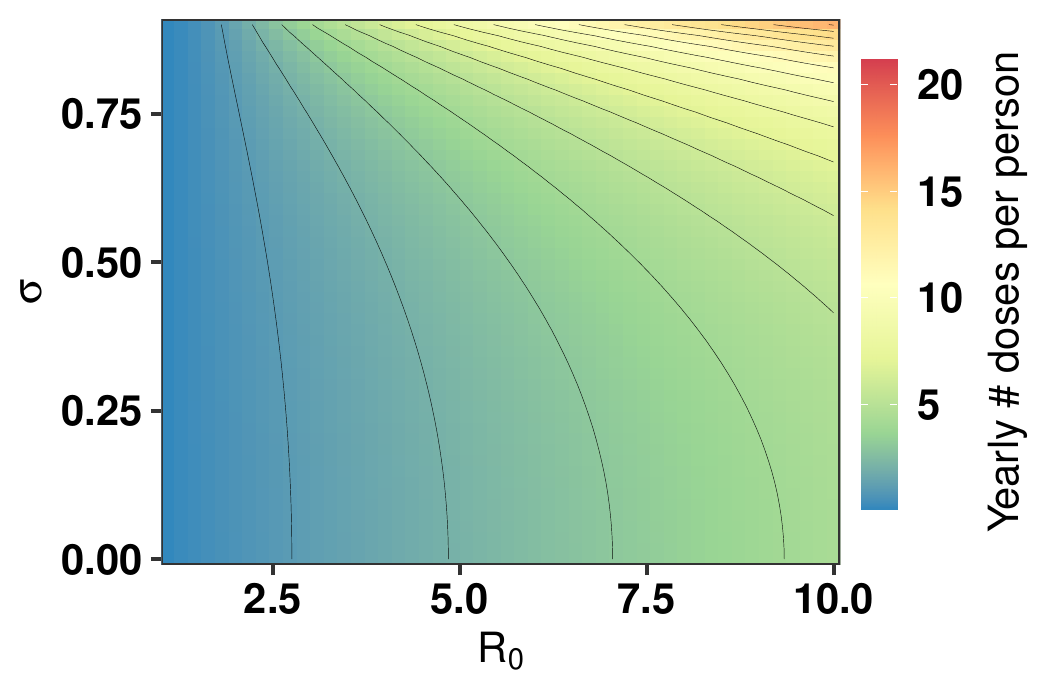}
        \caption{$p=50\%$.}
        \label{fig:critical vacc pply hSIRkS}
    \end{subfigure}
    \begin{subfigure}[b]{0.5\linewidth}        
        \centering
        \includegraphics[width=\linewidth]{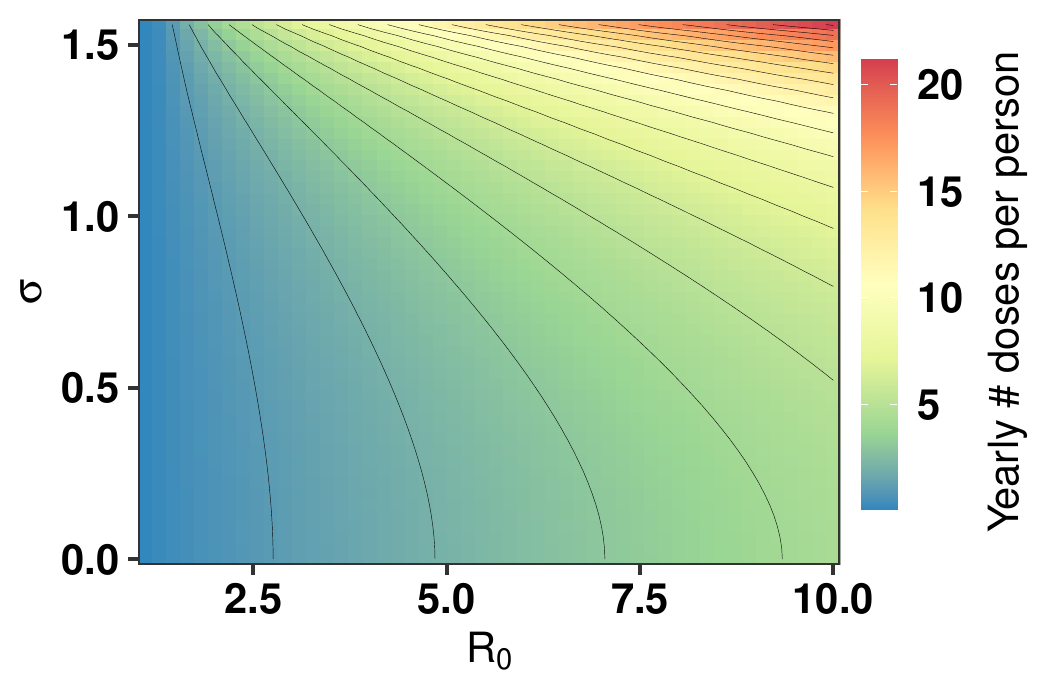}
        \caption{$p=75\%$.}
        \label{fig:ply hSIRkS}
    \end{subfigure}
\caption{Heatmaps of the critical vaccine supply for different values of $p$ in the uninformed scenario of the SIR\textsuperscript{($\infty$)}S model. The value at the origin is 0 and the contour interval is 1 yearly dose per person.}
\label{fig:crit_vacc_supply}
\end{figure}

Fig. \ref{fig:crit_vacc_supply} shows the minimum number of vaccine doses per person per year to achieve and maintain herd immunity according to the heterogeneous SIR\textsuperscript{$(\infty)$}S model. It is evident from the plots that the critical amount of vaccine supply in the continuous waning situation (for fixed $p$) is increasing in the coefficient of variation of population heterogeneity $\sigma$. Moreover, the bigger the fraction $p$ (of immune-weak type 1 individuals), the bigger the critical amount of vaccine supply. This indicates that heterogeneity in population immunity requires more frequent vaccination. It is worth mentioning that the optimal vaccine supply is not always increasing in heterogeneity when immunity wanes in one sudden leap as illustrated in Fig. \ref{fig:crit_vacc_supply_SIRS_app} in the Appendix. 

Table \ref{tab:critical vaccination frequency perfect} compares the critical vaccination frequency for different models when $R_0=5$ (and $\sigma=0.5$ for the heterogeneous models).  While the simple SIRS model suggests to vaccinate individuals every 15 months (0.81 doses per year) to maintain herd immunity, the heterogeneous exponentially waning immunity model increases this vaccination frequency to every $\approx 4.6$ months in the informed situation ($\approx 2.62$ doses per person per year), and to $\approx 4.4$ months in the more realistic uninformed situation ($\approx 2.76$ doses per person per year). Table \ref{tab:Re in theta} compares the value of $R_v$ for a given vaccine supply per year and shows that knowing individuals immunity status reduce the effective reproduction number compared to the uninformed situation, the difference is however moderate.

\begin{table}[ht]
	\renewcommand{\arraystretch}{1.5}
	\centering
 \caption{ Critical vaccination schemes for $R_0=5$ for the different models. The heterogeneous models are computed with $\sigma=0.5$ and $p=50\%$.}
\begin{tabular}{llp{2.5cm}}
\hline 
                   &   Vaccination frequency  (in months)    &   Yearly $\#$ doses per person   \\ \hline
            Hom.\ SIRS  &  15  &  0.81   \\
            Het.\ SIRS: informed  &  12.8 (10 / 17.7)~$^{1}$  &  0.94   \\
            Hom.\ exp.\ waning  &  5.5   &  2.15   \\
            Het. exp.\ waning: informed   &   4.6 (3.8 / 5.7) &  2.62   \\
            Het.\ exp.\ waning: uninformed   &  4.4  &  2.76   \\ \hline
            \multicolumn{3}{l}{
$^{1}$\footnotesize{x (y / z) means that vaccines are given to type-1 individuals every y time units and type-2 every z time units,}} \\
\multicolumn{3}{l}{
$~$\footnotesize{ resulting in vaccinating everyone every x time units on average.}}
\end{tabular}
\label{tab:critical vaccination frequency perfect}
\end{table}
\begin{table}[ht]
	\renewcommand{\arraystretch}{1.5}
	\centering
 \caption{Reproduction number for different values of individual vaccine supply per year given $R_0=5$ and $\sigma=0.5$ ($p=50\%$) under exponential waning of immunity.}
 \label{tab:Re in theta}
\begin{tabular}{cccccc}
\hline
                   Yearly $\#$ doses per person & $\theta=0$ &  $\theta=1/2$    &   $\theta=1$    &    $\theta=2$ & $\theta=3$ \\ \hline
            Informed  & 5 & 2.99  &  2.08  &  1.26 & 0.90 \\ 
            Uninformed   & 5 & 3.01  &  2.09  &  1.29 & 0.93 \\ \hline
\end{tabular}
\end{table}

\subsection{Optimal vaccination: Leaky vaccine}
Table \ref{tab:critical vaccination frequency imperfect} compares the optimal vaccination frequency in case of a leaky vaccine for the considered models. It is clear from the table that the optimal vaccination frequency increases as the protection $e$ becomes smaller. While herd immunity could be achieved with imperfect vaccines with relatively high efficacy when immunity wanes at once (e.g. by approximately administrating $80\%$-effective vaccines every 8.9 months on average -- Table \ref{tab:critical vaccination frequency imperfect}),  herd immunity under continuous waning would require very high vaccine efficacy and that in both homogeneous and heterogeneous situations.
\begin{table}[ht]
    \addtolength{\tabcolsep}{-3pt}
	\renewcommand{\arraystretch}{1.5}
	\centering
 \caption{ Critical vaccination frequency (in months)  for the different models when $R_0=5$ ($\sigma=0.5$ and $p=50\%$ in heterogeneous settings).}
\begin{tabular}{p{3.5cm}llll}
\hline 
            Vaccine efficacy         &   $e=100\%$ (perfect)    &   $e=95\%$ &   $e=90\%$ &   $e=80\%$  \\ \hline
            SIRS  &  15   &  14 & 13.2 & 11.8  \\
            Het. SIRS: informed  &  12.8 (10 / 17.7)~$^{1}$   &  11.7 (8.7 / 17.7) & 10.7 (7.7 / 17.7) & 8.9 (6 / 17.7)  \\
            Hom. exp. waning  &  5.5   &  4.5 & 3.3 & -- \\
            Het. exp. waning: informed   &   4.6 (3.8 / 5.7)  &  3.7 (2.7 / 4.8) &  2.6 (1.8 / 3.4) & -- \\
            Het. exp. waning: uninformed   &  4.4  &  3.5 & 2.5 & --  \\ \hline
            \multicolumn{5}{l}{
$^{1}$\footnotesize{12.8 (10 / 17.7) means that vaccines are given to type-1 individuals every 10 months and type-2 every 17.7 months,}} \\
\multicolumn{5}{l}{
$~$\footnotesize{ resulting in vaccinating individuals every 12.8 months on average.}}
\end{tabular}
\label{tab:critical vaccination frequency imperfect}
\end{table}

\subsection{Two extreme models comparison}
\begin{figure}[ht]
\centering
    \begin{subfigure}[b]{0.48\linewidth}        
        \centering
        \includegraphics[width=\linewidth]{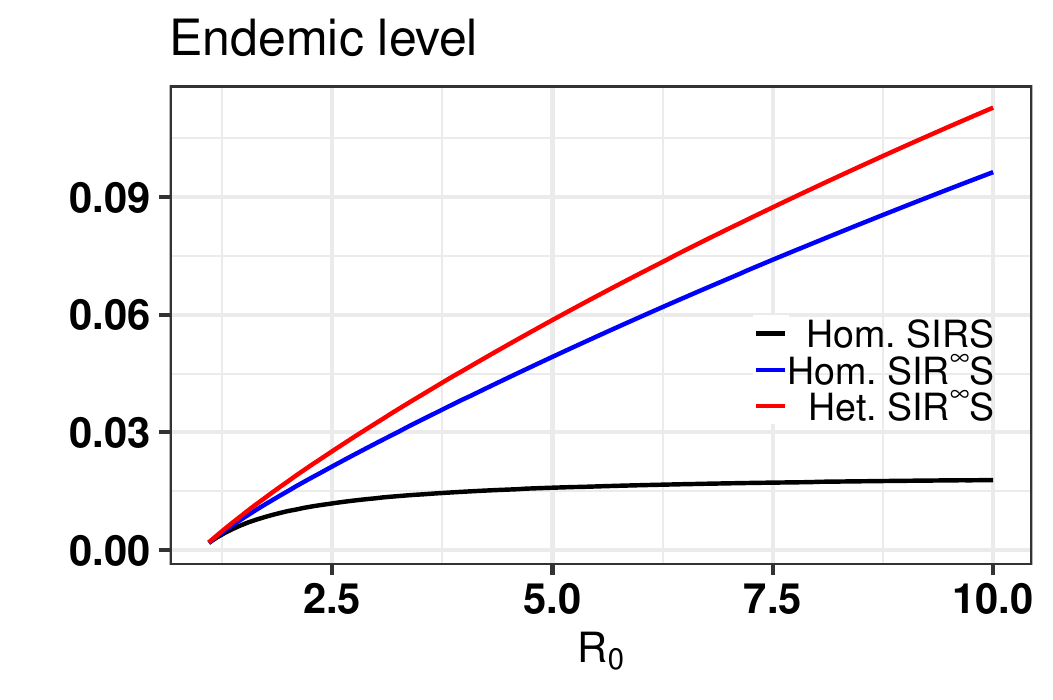}
        \caption{}
        \label{fig:magnitude_endemic}
    \end{subfigure}
    \begin{subfigure}[b]{0.48\linewidth}        
        \centering
        \includegraphics[width=\linewidth]{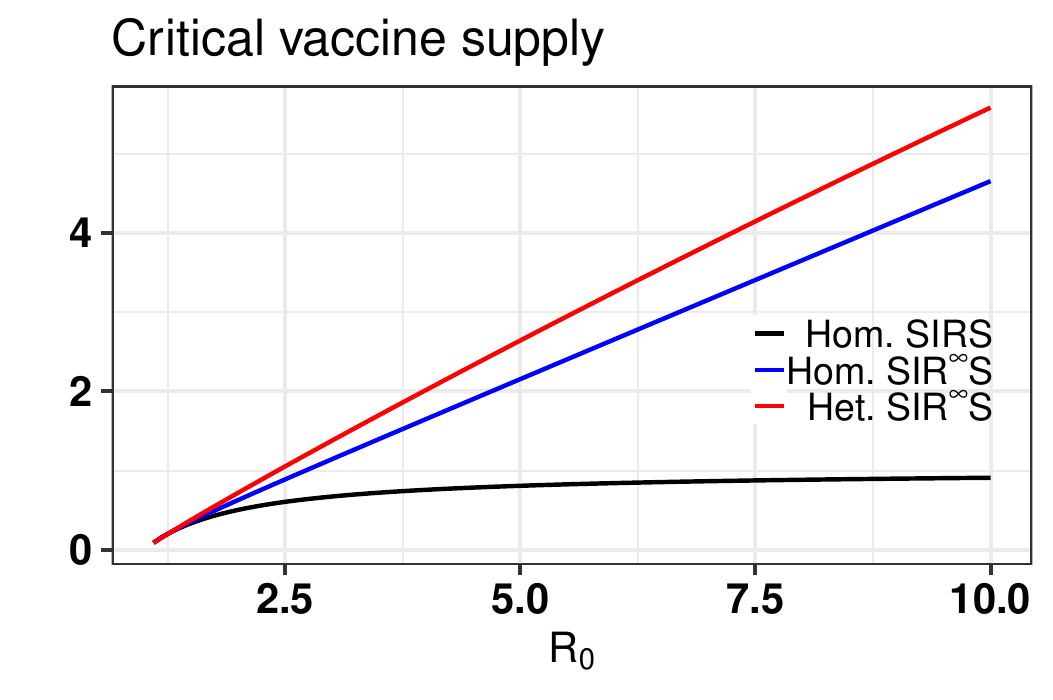}
        \caption{}
        \label{fig:magnitude_vacc}
    \end{subfigure}
    \caption{Comparison of the standard SIRS model, the homogeneous SIR\textsuperscript{$(\infty)$}S model, and the heterogeneous SIR\textsuperscript{$(\infty)$}S model with  $\sigma=0.5$ ($p=50\%$). (a) endemic levels and (b) the yearly number of vaccine doses per person required for herd immunity.}
    \label{fig:magnitude}
\end{figure}

In \cite{khalifi2022extending}, the standard SIRS model and the homogeneous SIR\textsuperscript{$(\infty)$}S model were compared and found that the latter has the larger endemic level and the higher critical vaccine supply. Fig. \ref{fig:magnitude} added a comparison with the heterogeneous SIR\textsuperscript{$(\infty)$}S model, with the critical vaccine supply plotted under the uninformed situation. It is clear that the biggest effect comes from the continuous waning of immunity compared to the sudden loss assumption. Still, heterogeneity makes the situation worse as it increases long-term prevalence and the critical vaccine coverage, in particular when heterogeneity is substantial.
\section{Discussion}\label{dis_sec}
In the current paper we have shown that if immunity wanes gradually but at different rates for different individuals, the effect of such heterogeneity is that endemic prevalence becomes higher, and when introducing vaccinations, more vaccines are required to reach and sustain herd immunity. This effect is shown to be substantial even when heterogeneity of immunity waning is moderate (e.g. coefficient of variation 0.5). An additional feature treated in our analysis is to distinguish between the informed situation where the waning heterogeneity is known and taken into account when designing vaccination policies, and the more likely uninformed scenario where such heterogeneities are unobserved. It is shown that the informed and uniformed situations differ in vaccination policies, but the required amount of vaccines for maintaining herd immunity is only moderately higher for the more likely uninformed situation.

This comparison between the homogeneous and heterogeneous situations, calibrated by assuming the same population average of cumulative immunity, hence has the heterogeneous situation as the worse case. As a consequence, models neglecting waning heterogeneities can estimate too low vaccination rates. This result is in contrast with many other comparisons in epidemic models in which the homogeneous situation is often the worst case scenario. Two such examples are \cite{ball1985deterministic} who considers variable susceptibility to the homogeneous situation where all individuals have the same susceptibility (see also \cite{elbasha2021vaccination}), and the second example is epidemics on networks where the final size is maximized when all individuals have equal degree (if the transmission rate is large enough) \cite{britton2012maximizing}.

Our new epidemic model with gradual waning rate with individual heterogeneity neglects many other factors affecting diseases dynamics. Such factors may for example include demographic structure, behaviour change of the population, elements of chance, individual heterogeneity also with respect to infectivity and susceptibility, social population structures, and so on. Here we neglect such aspects and focus on heterogeneity of immunity waning. It would be of interest to study the effect of waning heterogeneity also when including other realistic model features. It is our belief that the same qualitative observation remains: heterogeneity in immunity waning makes the situation worse, but clearly this needs to be shown.










\section*{Acknowledgement}
The authors are grateful to Gianpaolo Scalia-Tomba for helpful discussions.
\section*{Funding}
M.E.K. is grateful to NordForsk (project 105572) and T.B. is grateful to The Swedish Research Council (grant 2020-0474)  for financial support.
\section*{Ethics declarations}
The authors declare no competing interests.
\section*{Data availability}
All data generated or analysed during this study are included in this published article and its supplementary information files.
\bibliographystyle{vancouver}
\bibliography{mybib.bib}

\newpage
\appendix
\counterwithin{figure}{section}
\section{Appendix}\label{append: proofs of props}
\subsection{Proof}
\subsubsection{Proposition \ref{exist_uniq_het_SIRS}}\label{App_exist_uniq_het_SIRS}
First, it is easy to see that only the disease free equilibrium exists when $R_0<1$. Next, we assume that $R_0>1$ and let $\hat i=\hat i_1+\hat i_2$ to denote the endemic level. By equating the right hand side equations of \eqref{ODE1} to $0$ and after some simplifications, we get into 
\begin{align}\label{to_incr}
    \hat i =  \frac{ pR_0(\mu+\gamma) (\mu+\omega_1) \hat i }{\left(R_0(\mu+\gamma)\hat i + \mu\right)(\mu+\gamma+\omega_1) + \omega_1\gamma}+\frac{(1-p)R_0(\mu+\gamma)(\mu+\omega_2) \hat i }{\left(R_0(\mu+\gamma)\hat i + \mu\right)(\mu+\gamma+\omega_2) + \omega_2\gamma},
\end{align}
where we replaced $\beta$ by $R_0(\mu+\gamma)$.
That is, $\hat i$ is the (positive) fixed point of the function $\psi$ defined by
\begin{align*}
    \psi(x) = \frac{ pR_0(\mu+\gamma) (\mu+\omega_1) x }{\left(R_0(\mu+\gamma)x + \mu\right)(\mu+\gamma+\omega_1) + \omega_1\gamma}+\frac{(1-p)R_0(\mu+\gamma)(\mu+\omega_2) x }{\left(R_0(\mu+\gamma)x + \mu\right)(\mu+\gamma+\omega_2) + \omega_2\gamma},
\end{align*}
which is increasing on the positive real half line and verifies $\lim\limits_{x\to\infty} \psi(x)  <1$. Moreover, it can be shown that its derivative at $x=0$ satisfies $\psi'(0)>1$ as long as $R_0>1$ (and equals to 1 when $R_0=1$). Hence, $\psi$ has a unique positive fixed point $\hat i$, the endemic level, provided that $R_0>1$. Consequently, the equation \eqref{ODE1} has a unique endemic equilibrium if and only if $R_0>1$.
\subsubsection{Proposition \ref{end level increasing}}\label{App_end level increasing}
Now, we recall that $\omega_1=\omega/\left( 1-\sigma\sqrt{(1-p)/p}\right)$ and $\omega_2=\omega/\left( 1+\sigma\sqrt{p/(1-p)}\right)$ with $0\le\sigma<\sqrt{p/(1-p)}$.
To prove that the endemic level is increasing in $\sigma$, it is enough to show that the right hand side Eq. \eqref{to_incr} is increasing in $\sigma$.
This function could be defined by
\begin{align*}
    f(\sigma) = & \frac{ p(\mu+\omega) - \sigma \mu\sqrt{p(1-p)} }{\left(\beta i + \mu\right)(\mu+\gamma)+\omega\left(\beta i + \mu+\gamma\right) - \sigma\sqrt{(1-p)/p}\left(\beta i + \mu\right)(\mu+\gamma)}\\
    & + \frac{ (1-p)(\mu+\omega) + \sigma \mu\sqrt{p(1-p)} }{\left(\beta i + \mu\right)(\mu+\gamma)+\omega\left(\beta i + \mu+\gamma\right) + \sigma\sqrt{p/(1-p)}\left(\beta i + \mu\right)(\mu+\gamma)},
\end{align*}
Then, by differentiating, we obtain that the sign of $f'(\sigma)$ is the same as the sign of
\begin{align*}
     2 \left( \left(\beta i + \mu\right)(\mu+\gamma)+\omega\left(\beta i + \mu+\gamma\right) \right) + \sigma \left(\beta i + \mu\right)(\mu+\gamma) \left(\sqrt{\frac{p}{1-p}}-\sqrt{\frac{1-p}{p}} \right),
\end{align*}
which is positive provided that $p\ge 1/2$. Hence, $f$ is increasing in $\sigma$, and so is the endemic level.

\subsubsection{Proposition \ref{exist_uniq_het_contin}}\label{App_exist_uniq_het_contin}
From the infective equations in Eq. \eqref{ODE-PDE1}, and using the fact that $$s_1(t)+s_2(t)+\int_0^t r_1(\tau)\,d\tau+\int_0^t r_2(\tau)\,d\tau=1-(i_1(t)+i_2(t)),$$ the differential equation of the total infective fraction $i=i_1+i_2$ verifies
    \begin{align}\label{rhs}
        i'(t) \leq \beta i(t) \left( -i(t) + 1 - 1/R_0 \right).
    \end{align}
For any positive initial point $z_0$, the solution to the ODE $z'(t)=\beta z(t) \left( -z(t) + 1 - 1/R_0 \right)$ converges to 0 when $R_0\leq 1$. Then from the Ineq. \eqref{rhs}, we obtain that $i(t)\rightarrow0$ as $t\rightarrow\infty$ when $R_0\leq 1$. Hence, the infective and recovered fractions vanish. Moreover, the total susceptible fraction $s_1+s_2$ converges to 1. This proves the first assertion of the Proposition \ref{exist_uniq_het_contin}.
    
Now, we proceed to prove the second assertion of Proposition \ref{exist_uniq_het_contin}.
Solving the endemic equilibrium of system \eqref{ODE-PDE1} allows to write
\begin{align}\label{ODE-end-sol}
		&\frac{\mu}{2} - \beta s_l (i_1+i_2) - \mu s_l  = 0,\\
		 &\dfrac{\partial r_l(a)}{\partial a} = - \beta  \left( 1-e^{-\omega_l a}\right)\, r_l(a) (i_1+i_2) - \mu r_l(a),
   \quad r_l(0) = \gamma i_l,
\end{align}
for $l\in\{1,2\}$, coupled with 
\begin{align}
   i_1 &= p - s_1 - \int_0^\infty r_1(\tau)\,d\tau , \mbox{ and }
   i_2 = 1-p - s_2 - \int_0^\infty r_2(\tau)\,d\tau.
\end{align}
It is easy to see that an endemic equilibrium verifies both $i_1\neq 0$ and $i_2\neq 0$. Set $i=i_1+i_2$, then solving the ordinary differential equation for the recovered equations, we obtain
\begin{align}
    r_l(\tau)= \gamma i_l \, \phi_l(i), \qquad l\in\{1,2\},
\end{align}
where $\phi_l, l=1,2,$ are the functions defined by
\begin{align}
    \phi_l(x) = \int_0^\infty\exp\left(-\mu \tau - \beta x \int_0^\tau \left( 1-e^{-\omega_l a}\right) da \right) d\tau.
\end{align}
Then, we arrived to
\begin{align}
    i_1 &= p \left( 1 - \frac{\mu}{\beta i + \mu} - \gamma i_1 \, \phi_1(i) \right),\nonumber\\
    i_2 &= (1-p) \left( 1 -\frac{\mu}{\beta i + \mu}  - \gamma i_2 \, \phi_2(i) \right).
\end{align}
Re-arranging both equations allows to write $i_1$ and $i_2$ in terms of $i$ as
\begin{align}\label{i1 i2 in terms of i}
    i_1 &= \frac{\beta i}{\beta i + \mu} \frac{p}{   1+\gamma \phi_1(i) },\nonumber\\
    i_2 &= \frac{\beta i}{\beta i + \mu} \frac{1-p}{   1+\gamma \phi_2(i) } .
\end{align}
Taking the sum, it yields that
\begin{align}
    i = \frac{\beta i}{\beta i+\mu} \left( \frac{p}{   1+\gamma \phi_1(i) } + \frac{1-p}{ 1+\gamma \phi_2(i) } \right).
\end{align}
 As $i\neq0$, we cancel one $i$ and get to the following equation
\begin{align}\label{end_eq}
     i =  \left(\frac{p }{  1+\gamma \phi_1(i) } + \frac{1-p }{  1+\gamma \phi_2(i)} \right) - \frac{\mu}{\beta}.
\end{align}
The right-hand side of \eqref{end_eq} is increasing in $i$ and smaller than $1-\mu/\beta$. Moreover, it converges to $\mu(R_0-1)/\beta$ as $i\rightarrow0$. On the other hand, as $\phi_l,l=1,2,$ are convex functions, the right-hand side of \eqref{end_eq} is a concave down function. That is, Eq. \eqref{end_eq} has a unique positive solution when $R_0>1$ and no positive solution when $R_0\leq 1$. Since the right-hand side functions of \eqref{i1 i2 in terms of i} are increasing in $i$, the steady points $i_1$ and $i_2$ are defined from $i$ uniquely. This completes the proof.

\subsection{Formulation of the SIR\textsuperscript{$(k)$}S model with heterogeneity}\label{model_formulation}
Here we present the details of the determination of the immunity levels $  \{ 1- f_{l,j}\}_{j=1}^{k-1} $  and the immunity jumps rates $  \{ c_{l,j}\}_{j=1}^k $ in the model \eqref{SIRkS} in the main text for fixed $l\in \{1,2\}$. Let $  \{ f_{l,j}\}_{j=1}^{k-1} $ be an increasing sequence of elements of $(0,1)$. An $l$-individual recently recovered stays perfectly immune for an exponentially time with mean duration $1/c_{l,1}$, after that immunity drops to $1 - f_{l,1}$. Each $1- f_{l,j}$ immunity level lasts for an exponentially time with mean $1/c_{l,j+1}$ for $j=1,\cdots,k-1$. The susceptibility levels and the rates are chosen to fit the exponential waning with rate $w_l$ and to satisfy the constant average cumulative immunity equation
\begin{align}\label{cum_eq}
   \frac{1}{c_{l,1}} + \sum_{j=1}^{k-1} \left( 1- f_{l,j}\right) \frac{1}{c_{l,j+1}} = \frac{1}{\omega_l}.
\end{align}
We mention that there is no unique way to define the immunity jumps and the rates above, yet their choice would not affect the results for typically large $k$ as long as all jumps become small and rates large.
In the informed situation, we choose $f_{l,j}=j/k$ so the immunity jumps by $1/k$ each step. We then define the rates by
\begin{align}
    \frac{1}{c_{l,1}}&= - \frac{1}{w_l} \log \left( \frac{k-1}{k} +  \frac{x}{k}\right),\\
    \frac{1}{c_{l,j}}&=\frac{1}{w_l} \left( -\log \left( \frac{k-j}{k} +  \frac{x}{k}\right) + \log \left( \frac{k-j+1}{k} +  \frac{x}{k}\right)  \right),\quad j=2,\cdots,k,
\end{align}
where $x\in(0,1)$ solves the cumulative immunity equation \eqref{cum_eq}, that is, the equation
\begin{align}
    -  \log \left( \frac{k-1}{k} +  \frac{x}{k}\right) + \sum_{j=1}^{k-1} \frac{k-j}{k} \left( -\log \left( \frac{k-j}{k} +  \frac{x}{k}\right) + \log \left( \frac{k-j+1}{k} +  \frac{x}{k}\right)  \right) = 1.
\end{align}
In the uninformed situation we assume fixed (in $l$) immunity jump rates, that is $c_{1,j}=c_{2,j}=c_{j}$ and set
$c_{j}=k-j+1$. We define the susceptibility levels by 
\begin{align}
    f_{l,j} = 1 - \exp\left( -\omega_l (1+x_l) \sum_{n=1}^{j}\frac{1}{c_j}  \right),
\end{align}
with $x_l\in(0,1)$ is the solution to the the cumulative immunity equation \eqref{cum_eq}, that is, the equation
\begin{align}
   \sum_{j=0}^{k-1} \frac{1}{k-j} \exp\left( -\omega_l(1+x_l) \sum_{n=1}^{j}\frac{1}{c_j} \right) = \frac{1}{\omega_l}.
\end{align}

\subsection{Disease-free equilibria}\label{App_DFE}
\textbf{Informed situation:} The disease-free equilibrium $E_0^{inf} = \left(\hat{s}_{1},\hat{s}_{2}, \hat{r}_{1,0}, \hat{r}_{2,0}, \cdots, \hat{r}_{1,k-1}, \hat{r}_{2,k-1} \right)$ of the models \eqref{SIRkS vacc inf} is given by
\begin{align}
\hat{s}_{l} &= \frac{p_l \mu}{\mu+\eta_{1,k} \left( 1 - c_{l,k}^k A_k^l B_{k-1}^l  \right) }, \,l=1,2, \\
\hat{r}_{l,j} &= \eta_{l,k} A_k^l B_j^l \hat{s}_{l} , \,\,j=1,\cdots,k-1, \,l=1,2,\\
\hat{r}_{l,0} &=  \frac{1}{2} - \hat{s}_{l} - \sum\limits_{j=1}^{k-1} \hat{r}_{l,j}, \,l=1,2,
\end{align}
where $A_k^l= \left( \mu+c_{l,1}- \sum\limits_{j=1}^{k-1}\eta_{l,j} B_j^l \right)^{-1} $ and $B_j^l = \prod\limits_{n=1}^{j}\frac{c_{l,n}}{\mu+c_{l,n+1}+\eta_{l,n}}$, for $j=1,\cdots,k-1$ and $l=1,2$. 

\noindent\textbf{Uninformed situation:} The disease-free equilibrium $E_0^{uni} = \left(\hat{s}_{1},\hat{s}_{2}, \hat{r}_{1,0}, \hat{r}_{2,0}, \cdots, \hat{r}_{1,k-1}, \hat{r}_{2,k-1} \right)$ of the models \eqref{SIRkS vacc uninf} is given by
\begin{align}
\hat{s}_{l} &= \frac{p_l \mu}{\mu+\eta_{k} \left( 1 - c_{k}^k A_k B_{k-1}  \right) }, \,l=1,2, \\
\hat{r}_{l,j} &= \eta_{k} A_k B_j \hat{s}_{l} , \,\,j=1,\cdots,k-1, \,l=1,2,\\
\hat{r}_{l,0} &=  \frac{1}{2} - \hat{s}_{l} - \sum\limits_{j=1}^{k-1} \hat{r}_{l,j}, \,l=1,2,
\end{align}
where $A_k= \left( \mu+c_{1}- \sum\limits_{j=1}^{k-1}\eta_{j} B_j \right)^{-1} $ and $B_j = \prod\limits_{n=1}^{j}\frac{c_{n}}{\mu+c_{n+1}+\eta_{n}}$, for $j=1,\cdots,k-1$. 

\subsection[Endemic level: varying p]{Endemic level: varying $p$}
\begin{figure}[h]
\centering
    \begin{subfigure}[b]{0.48\textwidth}        
        \centering
        \includegraphics[width=\textwidth]{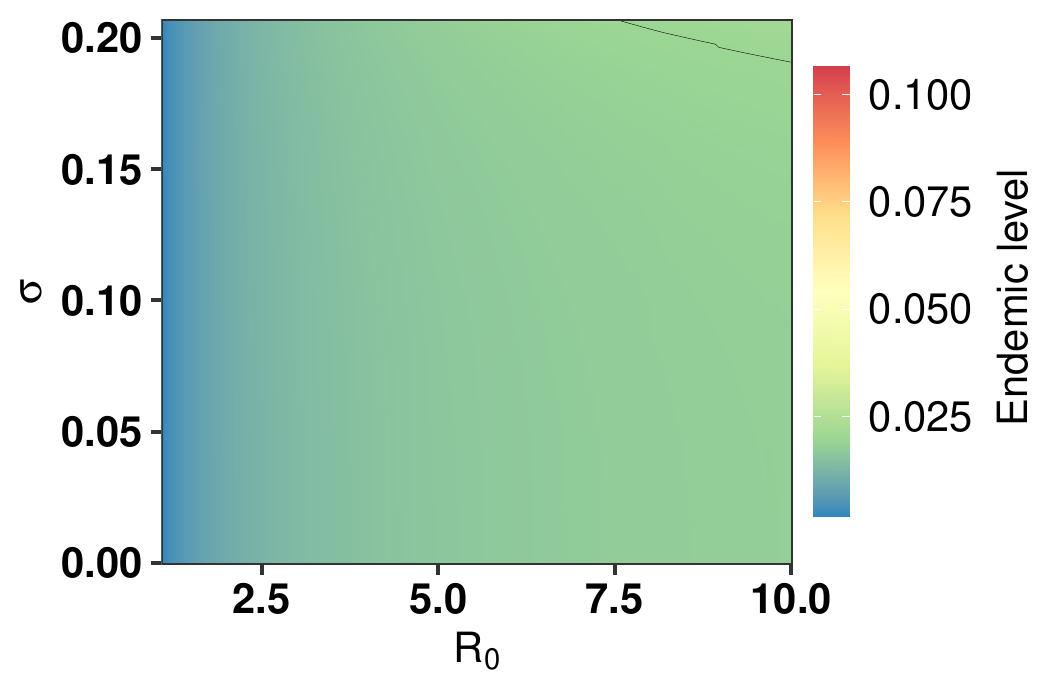}
        \caption{$p=5\%$.}
        \label{fig:endemic level hSIRS p05}
    \end{subfigure}
    \begin{subfigure}[b]{0.48\linewidth}        
        \centering
        \includegraphics[width=\linewidth]{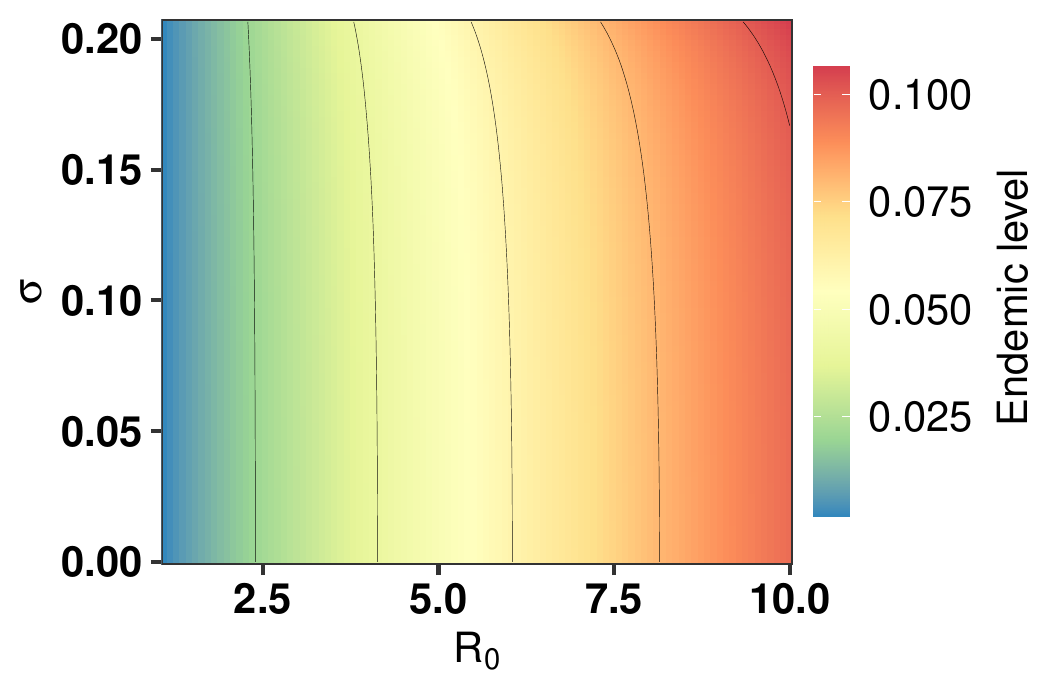}
        \caption{$p=5\%$.}
        \label{fig:endemic level hSIRkS p05}
    \end{subfigure}
    \begin{subfigure}[b]{0.48\linewidth}        
        \centering
        \includegraphics[width=\linewidth]{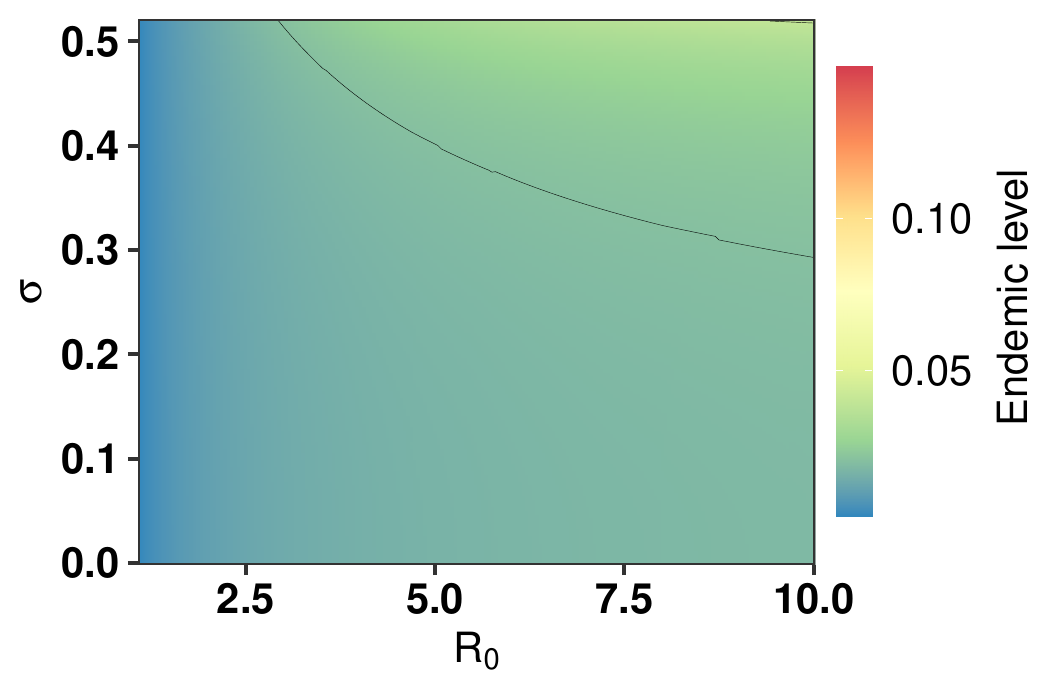}
        \caption{$p=25\%$.}
        \label{fig:endemic level hSIRS p25}
    \end{subfigure}
    \begin{subfigure}[b]{0.48\linewidth}        
        \centering
        \includegraphics[width=\linewidth]{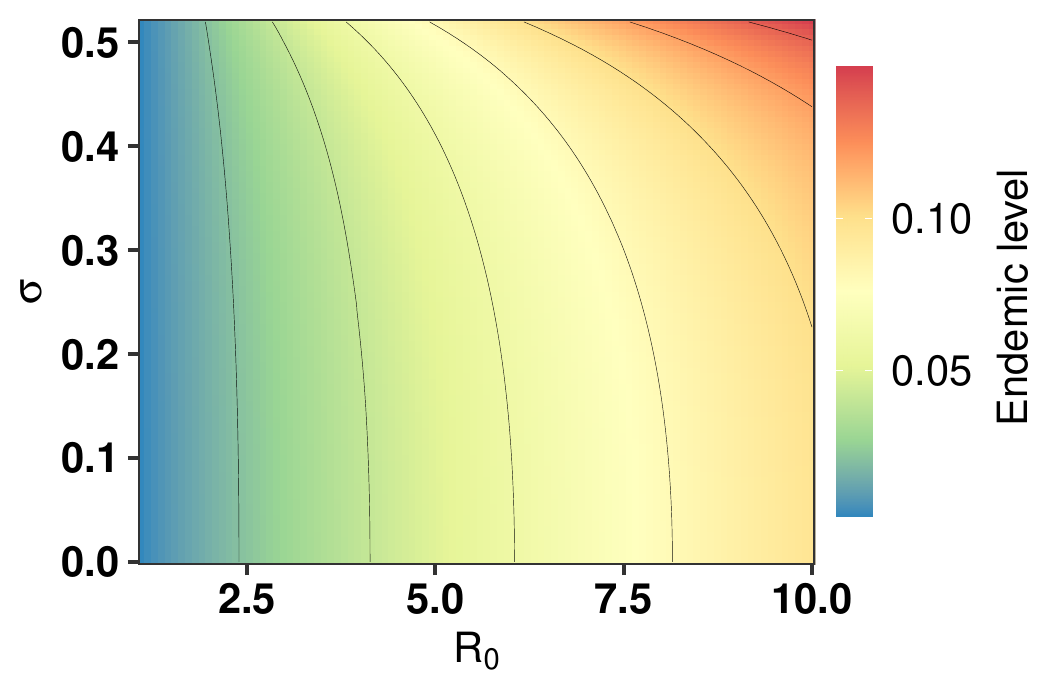}
        \caption{$p=25\%$.}
        \label{fig:endemic level hSIRkS p25}
    \end{subfigure}
    \begin{subfigure}[b]{0.48\linewidth}        
        \centering
        \includegraphics[width=\linewidth]{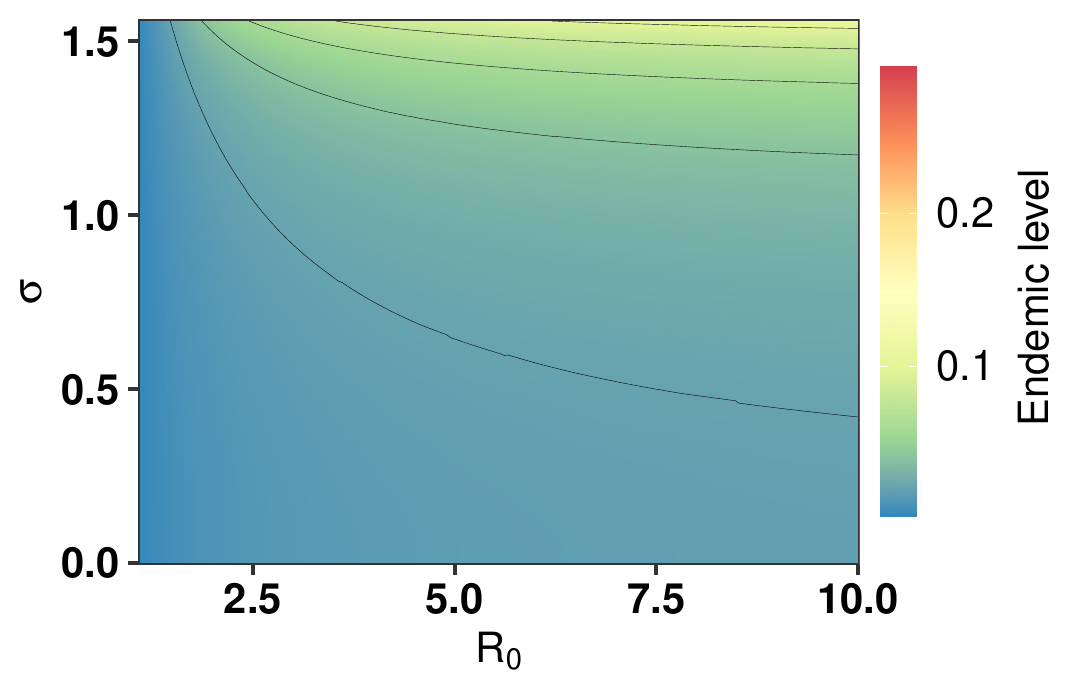}
        \caption{$p=75\%$.}
        \label{fig:endemic level hSIRS p75}
    \end{subfigure}
    \begin{subfigure}[b]{0.48\linewidth}        
        \centering
        \includegraphics[width=\linewidth]{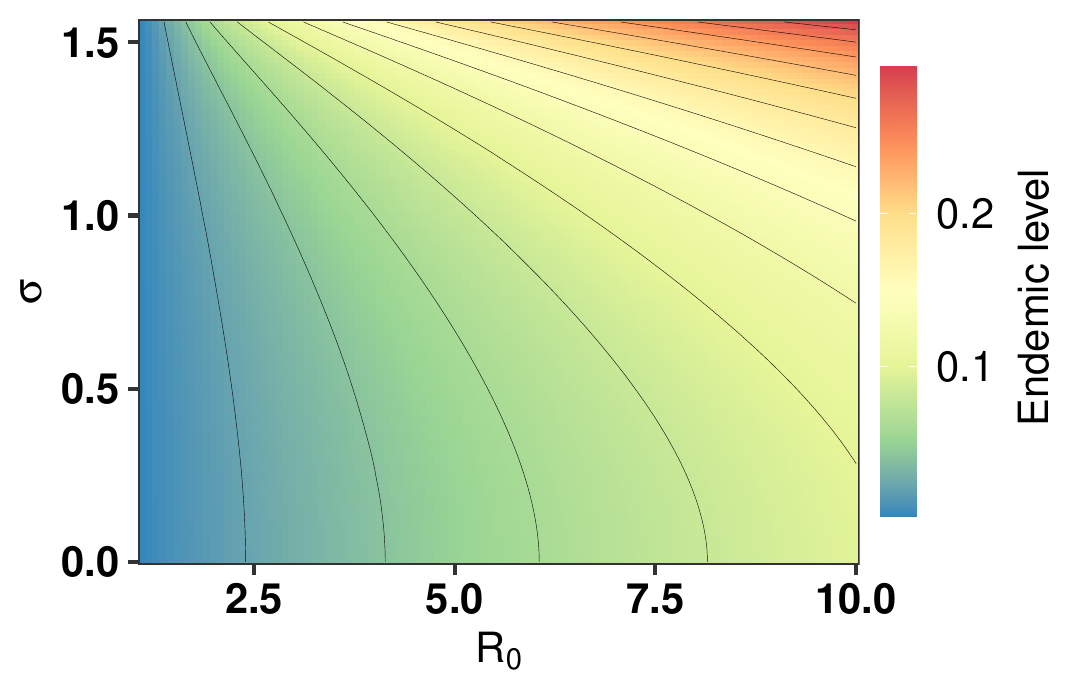}
        \caption{$p=75\%$.}
        \label{fig:endemic level hSIRkS p75}
    \end{subfigure}
    \begin{subfigure}[b]{0.48\linewidth}        
        \centering
        \includegraphics[width=\linewidth]{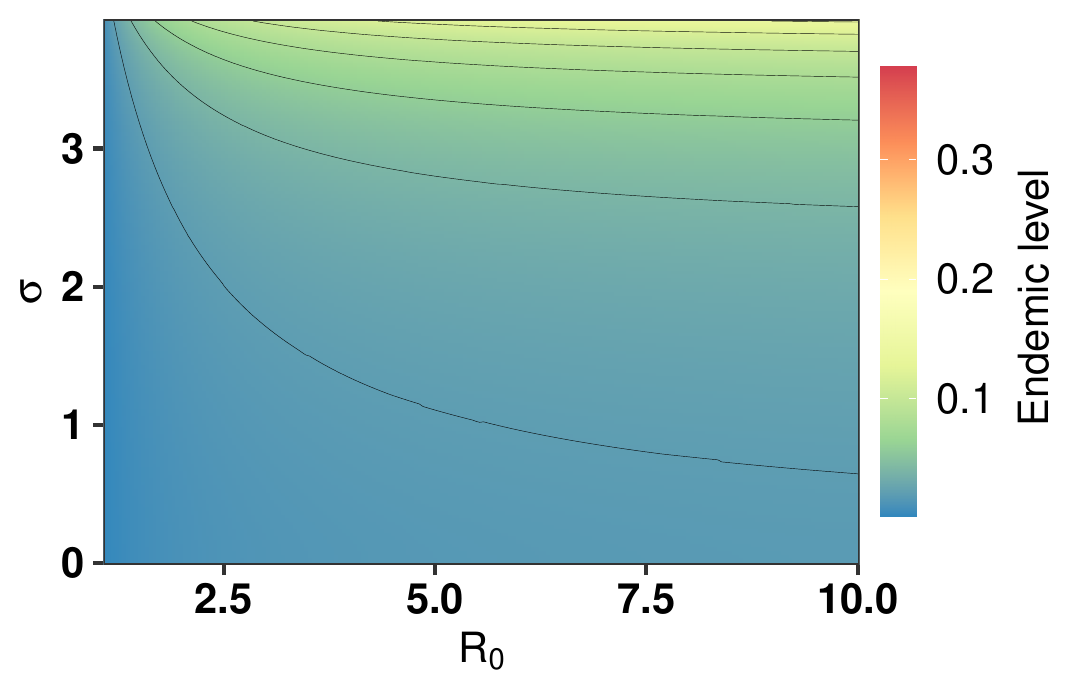}
        \caption{$p=95\%$.}
        \label{fig:endemic level hSIRS p95}
    \end{subfigure}
    \begin{subfigure}[b]{0.48\linewidth}        
        \centering
        \includegraphics[width=\linewidth]{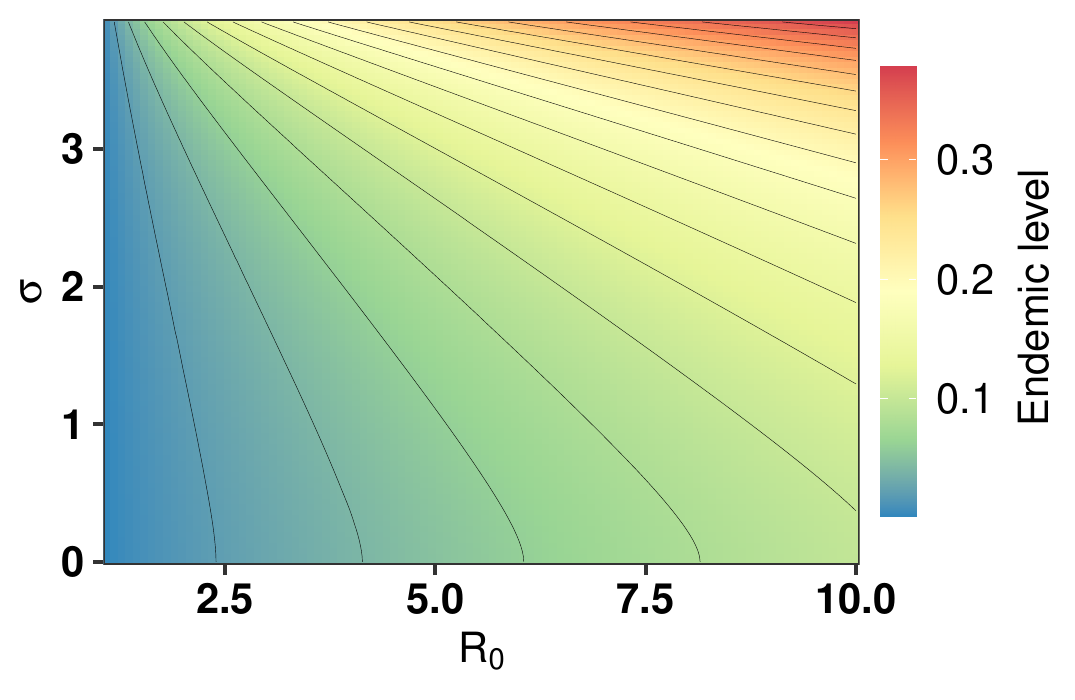}
        \caption{$p=95\%$.}
        \label{fig:endemic level hSIRkS p95}
    \end{subfigure}
    \caption{Heatmaps of the endemic level for different values of $R_0$ and $\sigma$. Left hand panel: Heterogeneous SIRS model. Right hand panel: Heterogeneous SIR\textsuperscript{$(\infty)$}S model. The value at the origin is 0 and the contour interval is 2\% of the population.}
    \label{fig:endemic levels p5-95_app}
\end{figure}

Fig. \ref{fig:endemic levels p5-95_app} shows how the endemic level varies with $R_0$ and $\sigma$ for different values of $p$ for the heterogeneous SIRS model (sudden loss of immunity) and the heterogeneous SIR\textsuperscript{$(\infty)$}S model (continuous waning).

\subsection{Critical vaccine supply when immunity wanes in one sudden leap}
\begin{figure}[!htb]
\centering
    \begin{subfigure}[b]{0.49\linewidth}        
        \centering
        \includegraphics[width=\linewidth]{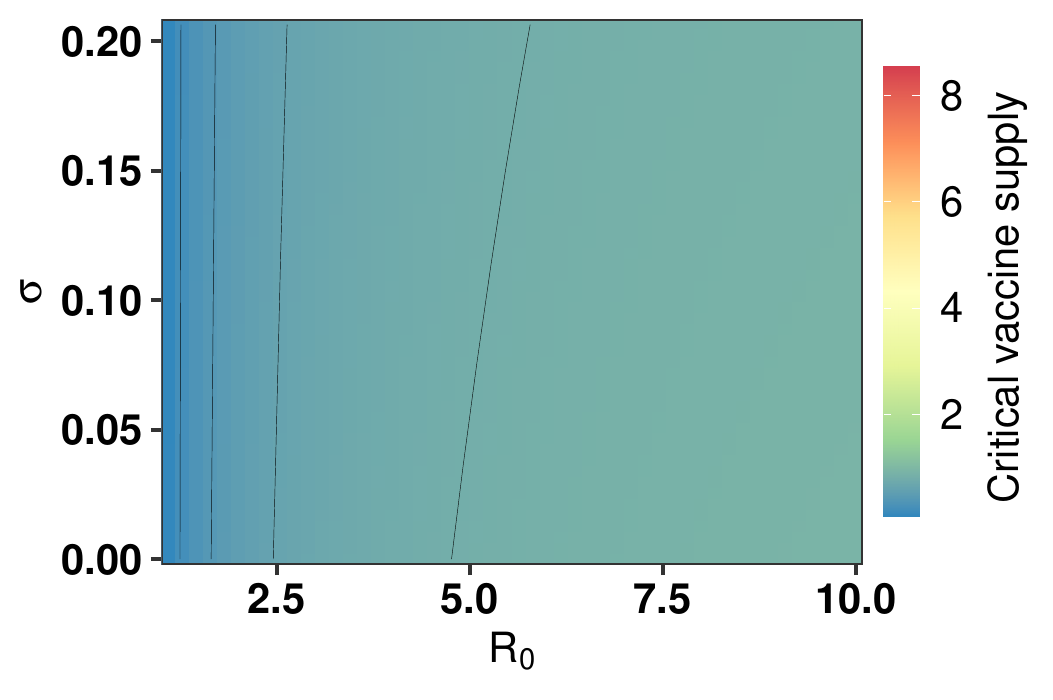}
        \caption{$p=5\%$.}
        \label{fig:het_SIRS_informed_p05}
    \end{subfigure}
    \begin{subfigure}[b]{0.49\linewidth}        
        \centering
        \includegraphics[width=\linewidth]{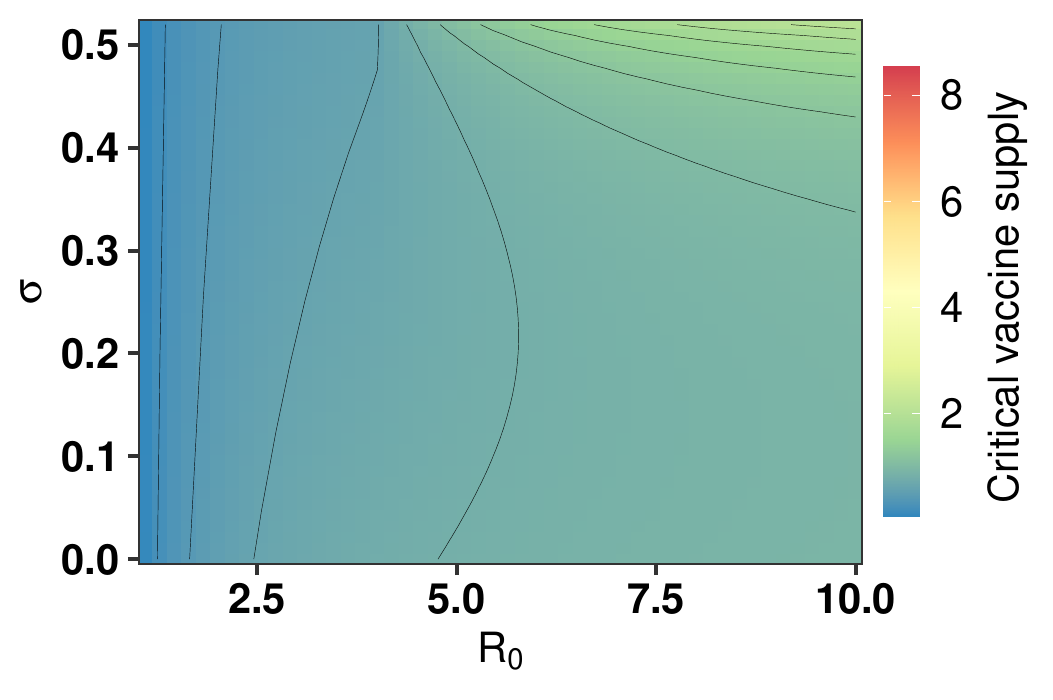}
        \caption{$p=25\%$.}
        \label{fig:het_SIRS_informed_p25}
    \end{subfigure}
    \begin{subfigure}[b]{0.49\linewidth}        
        \centering
        \includegraphics[width=\linewidth]{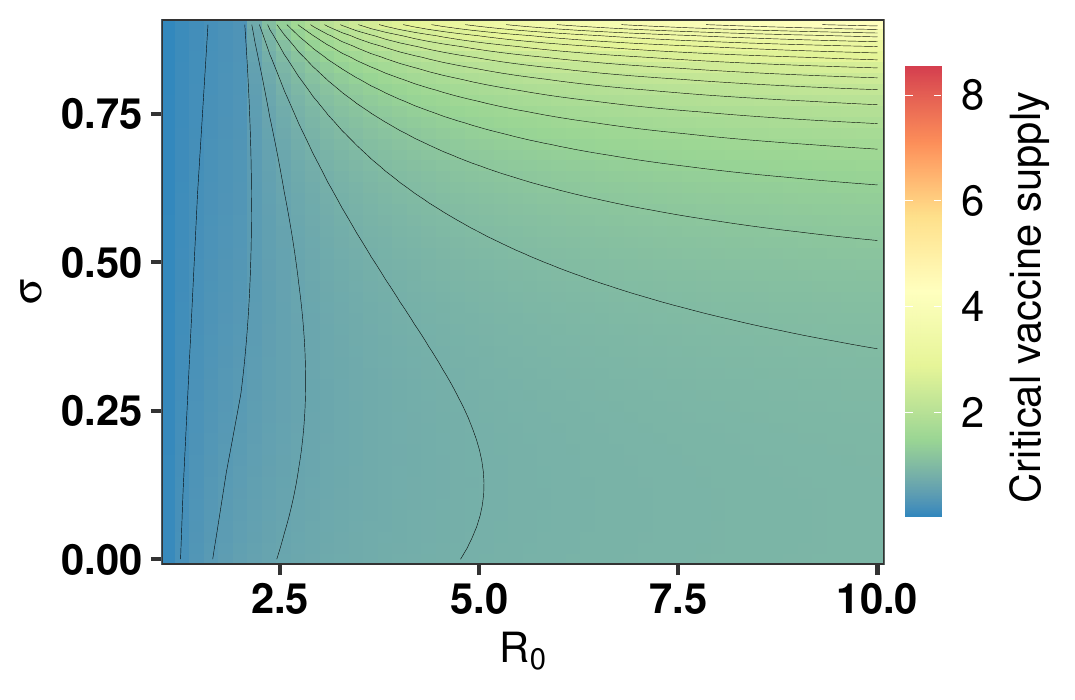}
        \caption{$p=50\%$.}
        \label{fig:het_SIRS_informed_p5}
    \end{subfigure}
    \begin{subfigure}[b]{0.49\linewidth}        
        \centering
        \includegraphics[width=\linewidth]{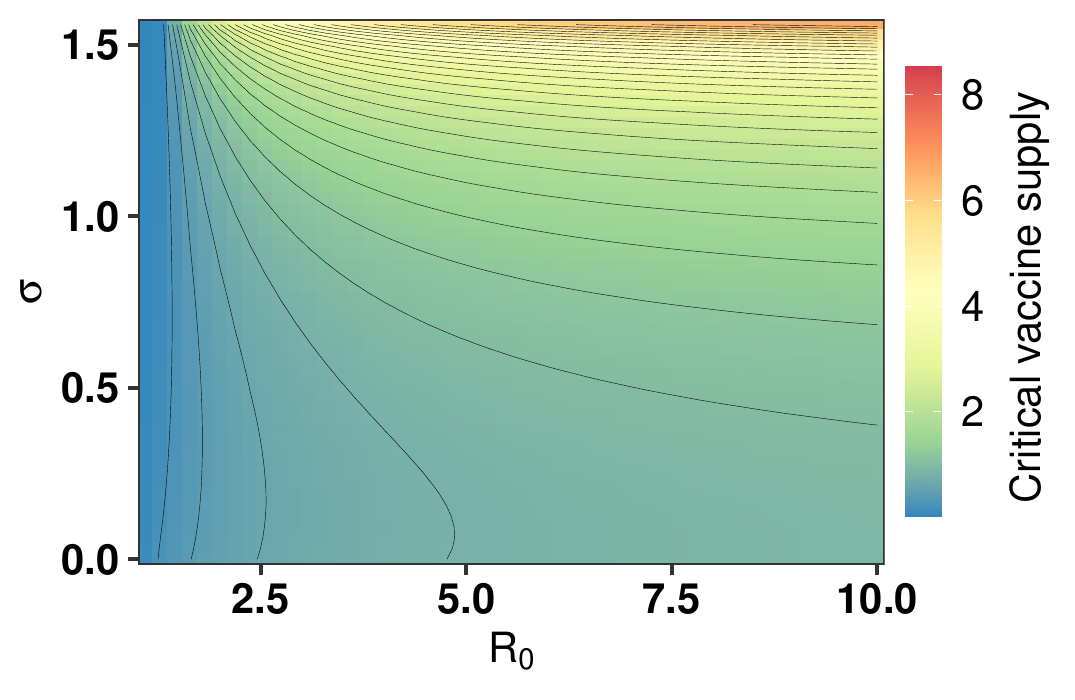}
        \caption{$p=75\%$.}
        \label{fig:het_SIRS_informed_p75}
    \end{subfigure}
    \begin{subfigure}[b]{0.5\linewidth}        
        \centering
        \includegraphics[width=\linewidth]{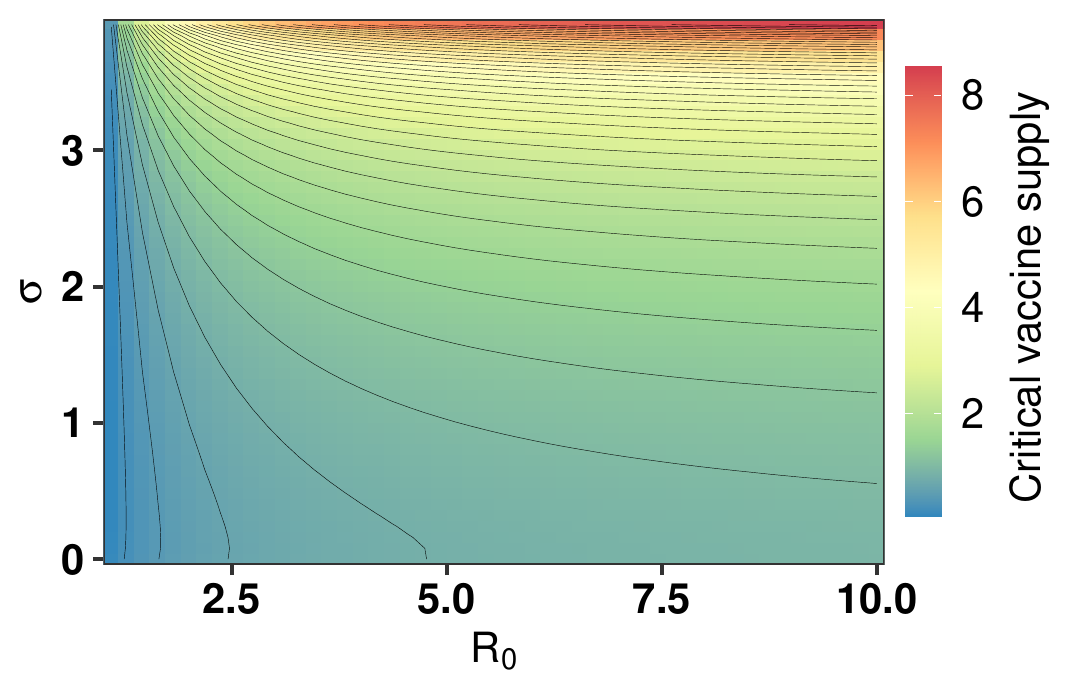}
        \caption{$p=95\%$.}
        \label{fig:het_SIRS_informed_p95}
    \end{subfigure}
\caption{Heatmaps of the critical vaccine supply for different values of $p$ in the informed SIRS model with sudden loss of immunity. The value at the origin is 0 and the contour interval is 1 yearly dose per person.}
\label{fig:crit_vacc_supply_SIRS_app}
\end{figure}

Fig. \ref{fig:crit_vacc_supply_SIRS_app} plots the critical vaccine supply for different values of $p$ for the informed case of the heterogeneous SIRS model with sudden loss of immunity.

\end{document}